\documentclass[twocolumn,english,aps,prb,floatfix]{revtex4}
\usepackage[T1]{fontenc}
\usepackage[latin2]{inputenc}
\usepackage{graphicx}
\usepackage{amssymb}

\makeatletter

\providecommand{\tabularnewline}{\\}

\usepackage{bm}

\usepackage{babel}
\makeatother
\begin{document}

\title{Formulae for zero-temperature conductance through a region with interaction}

\author{T. Rejec$^{1}$ and A. Ram\v{s}ak$^{1,2}$}

\affiliation{$^{1}$Jožef Stefan Institute, Jamova 39, SI-1000 Ljubljana, Slovenia\\
 $^{2}$Faculty of Mathematics and Physics, University of Ljubljana,
Jadranska 19, SI-1000 Ljubljana, Slovenia }

\begin{abstract}
The zero-temperature linear response conductance through an interacting
mesoscopic region attached to noninteracting leads is investigated.
We present a set of formulae expressing the conductance in terms of
the ground-state energy or persistent currents in an auxiliary system,
namely a ring threaded by a magnetic flux and containing the correlated
electron region. We first derive the conductance formulae for the
noninteracting case and then give arguments why the formalism is also
correct in the interacting case if the ground state of a system exhibits
Fermi liquid properties. We prove that in such systems, the ground-state
energy is a universal function of the magnetic flux, where the conductance
is the only parameter. The method is tested by comparing its predictions
with exact results and results of other methods for problems such
as the transport through single and double quantum dots containing
interacting electrons. The comparisons show an excellent quantitative
agreement. 
\end{abstract}
\maketitle

\section{introduction}

The measurements of the conductivity and the electron transport in
general are one of the most direct and sensitive probes in solid state
physics. In such measurements many interesting new phenomena were
signaled, in particular superconductivity, transport in metals with
embedded magnetic impurities and the related Kondo physics, heavy
fermion phenomena and the physics of the Mott-Hubbard transition regime.
In the last decade technological advances enabled controlled fabrication
of small regions connected to leads and \textit{\emph{the}} \textit{conductance,}
\textit{\emph{relating}} the current through such a region to the
voltage applied between the leads, also proved to be a relevant property
of such systems. There is a number of such examples, e.g. metallic
islands prepared by e-beam lithography or small metallic grains \cite{Delft01},
semiconductor quantum dots \cite{Kouwenhoven97}, or a single large
molecule such as a carbon nanotube or DNA. It is possible to break
a metallic contact and measure the transport properties of an atomic-size
bridge that forms in the break\cite{Agrait02}, or even measure the
conductance of a single hydrogen molecule, as reported recently in
Ref.~\cite{Smit02}. In all such systems, strong electron correlations
are expected to play an important role.

The transport in noninteracting mesoscopic systems is theoretically
well described in the framework of the Landauer-Büttiker formalism.
The conductance is determined with the Landauer-Büttiker formula~\cite{Landauer57,Landauer70,Buttiker86},
where the key quantity is the single particle transmission amplitude
$t(\varepsilon)$ for electrons in the vicinity of the Fermi energy.
The formula proved to be very useful and reliable, as long as electron-electron
interaction in a sample is negligible.

Although the Landauer-Büttiker formalism provides a general description
of the electron transport in noninteracting systems, it normally cannot
be used if the interaction between electrons plays an important role.
Several approaches have been developed to allow one to treat also
such systems. First of all, the Kubo formalism provides us with a
conductance formula which is applicable in the linear response regime
and has, for example, been used to calculate the conductance in Refs.~\cite{Oguri97a,Oguri01}.
A much more general approach was developed by Meir and Wingreen in
Ref.~\cite{Meir92}. Within the Keldysh formalism they manage to
express the conductance in terms of nonequilibrium Green's functions
for the sample part of the system. The formalism can be used to treat
systems at a finite source-drain voltage and can also be extended
to describe time-dependent transport phenomena \cite{Jauho94}. The
main theoretical challenge in these approaches is to calculate the
Green's function of a system. Except in some rare cases where exact
results are available, perturbative approaches or numerical renormalization
group studies are employed. 

In this paper we propose an alternative method for calculating the
conductance through such correlated systems. The method is applicable
only to a certain class of systems, namely to those exhibiting Fermi
liquid properties, at zero temperature and in the linear response
regime. However, in this quite restrictive domain of validity, the
method promises to be easier to use than the methods mentioned above.
We show that the ground-state energy of an auxiliary system, formed
by connecting the leads of the original system into a ring and threaded
by a magnetic flux, provides us with enough information to determine
the conductance. The main advantage of this method is the fact that
it is often much easier to calculate the ground-state energy (for
example, using variational methods) than the Green's function, which
is needed in the Kubo and Keldysh approaches. The conductance of a
Hubbard chain connected to leads was studied recently using a special
case of our method and DMRG \cite{Favand98,Molina02} and a special
case of our approach was applied in the Hartree-Fock analysis of anomalies
in the conductance of quantum point contacts \cite{Sushkov01}. The
method is related to the study of the charge stiffness and persistent
currents in one-dimensional systems \cite{Cheung88,Gogolin94,Aligia02,Prelovsek01}. 

The paper is organized as follows. In Section II we present the model
Hamiltonian for which the method is applicable. In Section III we
derive general formulae for the zero-temperature conductance through
a mesoscopic region with noninteracting electrons connected to leads.
In Section IV we extend the formalism to the case of interacting electrons.
We give arguments why the formalism is correct as long as the ground
state of the system exhibits Fermi liquid properties. In Section V
convergence tests for a typical noninteracting system are first presented.
Then we support our formalism also with numerical results for the
conductance of some non-trivial problems, such as the transport through
single and double quantum dots containing interacting electrons and
connected to noninteracting leads. These comparisons, including the
comparison with the exact results for the Anderson model, demonstrate
a good quantitative agreement. After the conclusions in Section VI
we present some more technical details in Appendix A. In Appendix
B we describe the numerical method used in Section V.

\section{Model Hamiltonian}

\begin{figure}[htbp]
\begin{center}\includegraphics[%
  width=8cm,
  keepaspectratio]{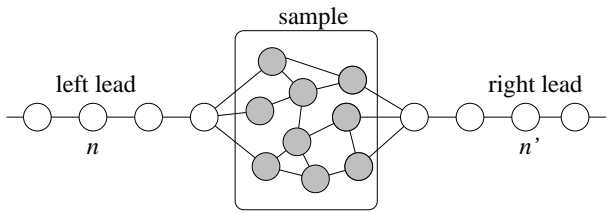}\end{center}

\caption{\label{cap:System}Schematic picture of the system described by Hamiltonian
(\ref{eq:Hamiltonian}).}
\end{figure}

In this Section we introduce a general Hamiltonian describing a mesoscopic
sample coupled to leads as shown in Fig.~\ref{cap:System}. The Hamiltonian
is a generalization of the well known Anderson impurity model \cite{Anderson61}.
We split the Hamiltonian into five pieces

\begin{equation}
H=H_{L}+V_{L}+H_{C}+V_{R}+H_{R},\label{eq:Hamiltonian}\end{equation}
where $H_{C}$ models the central region, $H_{L}$ and $H_{R}$ describe
the left and the right lead, and $V_{L}$ and $V_{R}$ are the tunneling
couplings between the leads and the central region. We can also split
the Hamiltonian into a term $H^{\left(0\right)}$ describing independent
electrons and a term $U$ describing the Coulomb interaction between
them \begin{equation}
H=H^{\left(0\right)}+U.\label{eq:H0plusU}\end{equation}
One can often neglect the interaction in the leads and between the
sample and the leads. We assume this is the case. Then the central
region is the only part of the system where one must take the interaction
into account

\begin{equation}
H_{\mathrm{C}}=H_{C}^{\left(0\right)}+U.\label{eq:HC}\end{equation}
Here $H_{C}^{\left(0\right)}$ describes a set of noninteracting levels\begin{equation}
H_{C}^{\left(0\right)}=\sum_{{{i,j\in C\atop \sigma}}}H_{Cji}^{\left(0\right)}d_{j\sigma}^{\dagger}d_{i\sigma},\label{eq:HCni}\end{equation}
where $d_{i\sigma}^{\dagger}$ ($d_{i\sigma}$) creates (destroys)
an electron with spin $\sigma$ in the state $i$. The states introduced
here can have various physical meanings. They could represent the
true single-electron states of the sample, for example different energy
levels of a multi-level quantum dot or a molecule. In this case, the
matrix $H_{Cji}^{\left(0\right)}$ is diagonal and its elements are
the single-electron energies of the system. Another possible interpretation
of Hamiltonian (\ref{eq:HCni}) is that the states $i$ are local
orbitals at different sites of the system. In this case, the diagonal
matrix elements of $H_{C}^{\left(0\right)}$ are the on-site energies
for these sites, while the off-diagonal matrix elements describe the
coupling between different sites of the system. The sites could have
a direct physical interpretation, such as dots in a double quantum
dot system or atoms in a molecule, or they could represent fictitious
sites obtained by discretization of a continuous system. There are
other possible choices of basis states for the central region. For
example, in a system consisting of two multi-level quantum dots one
could use single-electron basis states for each of the dots and describe
the coupling between the dots with tunneling matrix elements. 

The Coulomb interaction between electrons in the sample is given by
an extended Hubbard-type coupling\begin{equation}
U=\frac{1}{2}\sum_{{{i,j\in C\atop \sigma,\sigma^{\prime}}}}U_{ji}^{\sigma\sigma^{\prime}}n_{j\sigma}n_{i\sigma^{\prime}},\label{eq:HU}\end{equation}
where $n_{i\sigma}=d_{i\sigma}^{\dagger}d_{i\sigma}$ is the operator
counting the number of electrons with spin $\sigma$ at site $i$.
For convenience, we wrote down only the expression for the Coulomb
interaction in the case, where basis states represent different sites
in real space. The expression becomes more complicated if a more general
basis set is used. 

We describe the leads or contacts as two semi-infinite, tight-binding
chains 

\begin{equation}
H_{\mathrm{L}\left(R\right)}=-t_{0}\sum_{{{i,i+1\in L\left(R\right)\atop \sigma}}}c_{i\sigma}^{\dagger}c_{i+1\sigma}+\mathrm{h.c.},\end{equation}
where $c_{i\sigma}^{\dagger}$ ($c_{i\sigma}$) creates (destroys)
an electron with spin $\sigma$ on site $i$ and $t_{0}$ is the hopping
matrix element between neighboring sites. Such a model adequately,
at least for energies low or comparable to $t_{0}$, describes a noninteracting,
single-mode and homogeneous lead. It would be easy to generalize the
lead Hamiltonian to describe a more realistic system, for example
by modeling the true geometry or allowing for a self-consistent potential
due to interaction between electrons. However, the physics we are
interested in, is usually not changed dramatically by not including
these details into the model Hamiltonian and therefore, we will not
discuss this issue into detail.

Finally, there is a term describing the coupling between the sample
and the leads,

\begin{equation}
V_{L\left(R\right)}=\sum_{{{j\in L\left(R\right)\atop {{i\in C\atop \sigma}}}}}V_{L\left(R\right)ji}c_{j\sigma}^{\dagger}d_{i\sigma}+\mathrm{h}.\mathrm{c}.,\label{eq:Hhop}\end{equation}
where $V_{L\left(R\right)ji}$ is the hopping matrix element between
state $i$ in the sample and site $j$ in a lead. 

\begin{figure}[htbp]
\begin{center}\includegraphics[%
  width=5.6cm,
  keepaspectratio]{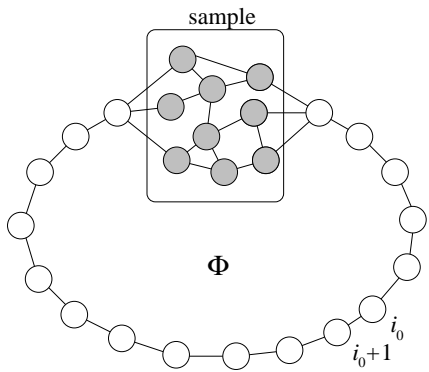}\end{center}

\caption{\label{cap:System1}The sample embedded in a ring formed by joining
the left and right leads of the system in Fig.~\ref{cap:System}.
Magnetic flux $\Phi$ penetrates the ring. }
\end{figure}

In the following Sections we discuss the conductance through the system
introduced above. To derive the conductance formulae, we will need
a slightly modified system. This auxiliary system is a ring formed
by connecting the ends of the left and right leads of the original
system as shown in Fig.~\ref{cap:System1}. The ring is threaded
by a magnetic flux $\Phi$ in such a way that there is no magnetic
field in the region where electrons move. We can then perform the
standard Peierls substitution \cite{Peierls33} and transform the
hopping matrix elements of the Hamiltonian (\ref{eq:Hamiltonian})
according to \begin{equation}
t_{ji}\rightarrow t_{ji}e^{i\frac{e}{\hbar}\int_{\mathbf{x}_{i}}^{\mathbf{x}_{j}}\mathbf{A}\cdot\mathrm{d}\mathbf{x}},\label{tij}\end{equation}
where $\mathbf{x}_{i}$ is the position of site $i$ and $\mathbf{A}$
is the vector potential due to the flux, obeying 

\begin{equation}
\Phi=\frac{\hbar}{e}\phi=\oint\mathbf{A}\cdot\mathrm{d}\mathbf{x}.\label{condition}\end{equation}
Here we defined a dimensionless magnetic flux $\phi$. The energy
of the system is periodic in $\phi$ with a period of $2\pi$ and
depends only on value of $\phi$ and not on any details of how the
flux is produced. If the original Hamiltonian (\ref{eq:Hamiltonian})
obeys the time-reversal symmetry, the energy does not change if the
magnetic field is reversed, \begin{equation}
E\left(-\phi\right)=E\left(\phi\right).\label{ereverse}\end{equation}

\section{Conductance of a noninteracting system\label{sec:sec23}}

In this Section we limit the discussion to noninteracting systems,
i.e. we set $U=0$ in Eq.~(\ref{eq:H0plusU}). In such systems, the
Landauer-Büttiker formula\cite{Landauer57,Landauer70,Buttiker86}\begin{equation}
G=G_{0}\left|t\left(\varepsilon_{F}\right)\right|^{2},\label{eq:landauer}\end{equation}
 which relates the zero-temperature conductance $G$ to the transmission
probability $\left|t\left(\varepsilon_{F}\right)\right|^{2}$ for
electrons at the Fermi energy $\epsilon_{F}$, can be applied. The
proportionality coefficient, $G_{0}=\frac{2e^{2}}{h}$, is the quantum
of conductance. Below we first derive a set of formulae, which relate
the transmission probability, and consequently the conductance, to
single-electron energy levels of the auxiliary ring system introduced
in the previous Section. Then we derive another set of formulae, relating
the conductance to the ground-state energy of the auxiliary system.
One of these formulae was derived before in Ref.~\cite{Sushkov01},
and a limiting case of another one was discussed in Refs.~\cite{Favand98,Molina02}.
Here we present a unified approach to the problem, from which these
results emerge as special cases.

\subsection{Formulae relating conductance to single-electron energy levels\label{sub:FormulaeSE}}

\begin{figure}[htbp]
\begin{center}\includegraphics[%
  width=8.5cm,
  keepaspectratio]{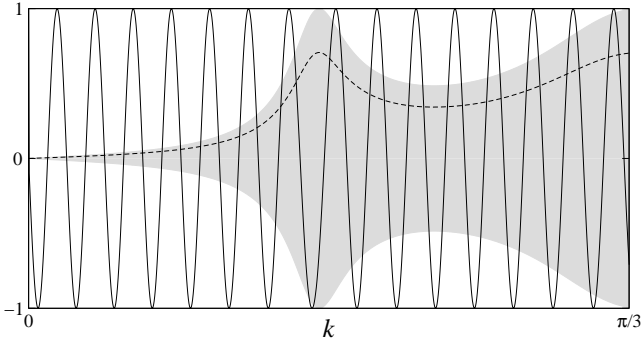}\end{center}

\caption{\label{cap:figcos}A graphical representation of the eigenvalue equation
(\ref{eigen}). The shaded region represents the allowed values of
the left hand side of the equation for different values of magnetic
flux (for example, the dashed line shows the values for $\phi=\frac{\pi}{4}$).
The full line represents the right hand side of the eigenvalue equation.
The system is presented in Fig.~\ref{cap:testnisys}, $N=100$.}
\end{figure}

Let us consider eigenstates of an electron moving on a ring system
introduced in the previous Section. We will be interested only in
energies of these states and not in the precise form of wavefunctions.
The energy of an electron on a ring penetrated by a magnetic flux
$\phi$ depends only on the magnitude of the flux and therefore, any
vector potential fulfilling condition (\ref{condition}) is good for
our purpose. We choose a vector potential constant everywhere except
between sites $i_{0}$ and $i_{0}+1$, both in the lead part of the
ring as shown in Fig.~\ref{cap:System1}. The hopping matrix element
between the two sites is thus modified to $t_{0}e^{i\phi}$. With
no flux penetrating the ring, the electron's wavefunction in the lead
part of the system is $ae^{iki}+be^{-iki}$, where $k$ is the electron's
wavevector and $a$ and $b$ are amplitudes determined by properties
of the central region.  If there is a flux through the ring, the wavefunction
is modified. The Schrödinger equations for sites $i_{0}$ and $i_{0}+1$
show us that the appropriate form is $ae^{iki}+be^{-iki}$ for $i\leq i_{0}$
and $ae^{-i\phi}e^{iki}+be^{-i\phi}e^{-iki}$ for $i>i_{0}$. The
scattering matrix of the central region provides a relation between
coefficients $a$ and $b$,\begin{equation}
\left(\begin{array}{c}
be^{-i\phi}e^{ikN}\\
a\end{array}\right)=\left(\begin{array}{cc}
r_{k} & t_{k}^{\prime}\\
t_{k} & r_{k}^{\prime}\end{array}\right)\left(\begin{array}{c}
ae^{-i\phi}e^{-ikN}\\
b\end{array}\right).\label{eq:linsyst}\end{equation}
The elements of the scattering matrix, $t_{k}$ and $r_{k}$ ($t_{k}^{\prime}$
and $r_{k}^{\prime}$), are the transmission and reflection amplitudes
for electrons coming from the left (right) lead, and $N$ is the number
of sites in the lead part of the ring. We added phase factors $e^{\pm ikN}$
to the {}``left lead'' amplitudes to compensate for the phase difference
an electron accumulates as it travels through the lead part of the
ring. The scattering matrix defined this way does not depend on $N$
and $\phi$, and equals the scattering matrix of the original, two-lead
system. Eq.~(\ref{eq:linsyst}) is a homogeneous system of linear
equations, solvable only if the determinant is zero. Using the unitarity
property of the scattering matrix, the eigenenergy condition becomes\begin{equation}
t_{k}^{\prime}e^{i\phi}+t_{k}e^{-i\phi}=e^{ikN}+\frac{t_{k}}{t_{k}^{\prime\ast}}e^{-ikN}.\end{equation}
 We assume the Hamiltonian of the original, two lead system obeys
the time-reversal symmetry and therefore, the scattering matrix is
symmetric~\cite{Datta95}, $t_{k}=t_{k}^{\prime}$. Expressing the
transmission amplitude in terms of its absolute value and the scattering
phase shift $t_{k}=\left|t_{k}\right|e^{i\varphi_{k}}$, we arrive
at the final form of the eigenenergy equation

\begin{equation}
\left|t_{k}\right|\cos\phi=\cos\left(kN-\varphi_{k}\right).\label{eigen}\end{equation}
In Fig.~\ref{cap:figcos} a graphical representation of this equation
is presented.

To extract the transmission probability $\left|t_{k}\right|^{2}$,
we proceed by differentiating the eigenvalue equation with respect
to $\cos\phi$\begin{eqnarray}
 &  & \frac{\partial\left|t_{k}\right|}{\partial\cos\phi}\cos\phi+\left|t_{k}\right|=\nonumber \\
 &  & \quad=-\sin\left(kN+\varphi_{k}\right)\left(N\frac{\partial k}{\partial\cos\phi}+\frac{\partial\varphi_{k}}{\partial\cos\phi}\right)=\nonumber \\
 &  & \quad=\pm\sqrt{1-\left|t_{k}\right|^{2}\cos^{2}\phi}\left(N\frac{\partial k}{\partial\cos\phi}+\frac{\partial\varphi_{k}}{\partial\cos\phi}\right).\quad\quad\label{eq:eqt1}\end{eqnarray}
 The sign of the last expression depends on weather $k$ belongs to
a decreasing ($+$) or an increasing ($-$) branch of the cosine function
in Eq.~(\ref{eigen}), or equivalently, if we are interested in an
eigenstate with odd ($+$) or even ($-$) $n$, where $n$ indexes
the eigenstates from the one with the lowest energy upward. Let us
choose an eigenstate and consider how the corresponding wavevector
$k$ changes as the magnetic flux $\phi$ is varied from 0 to $\pi$.
It is evident that the variation in $k$ is of the order of $\frac{1}{N}$
as the cosine function in the right hand side of Eq.~(\ref{eigen})
oscillates with such a period. Let as assume that the number of sites
in the ring is large enough that transmission amplitude does not change
appreciably in this interval\begin{equation}
\left|\frac{\partial t_{k}}{\partial k}\right|\frac{\pi}{N}\ll1.\label{cond1}\end{equation}
 Then the derivatives $\frac{\partial k}{\partial\cos\phi}$, $\frac{\partial\left|t_{k}\right|}{\partial\cos\phi}$
and $\frac{\partial\varphi_{k}}{\partial\cos\phi}$ are of the order
of $\frac{1}{N}$ and Eq.~(\ref{eq:eqt1}) simplifies to\begin{equation}
\left|t_{k}\right|=\pm\sqrt{1-\left|t_{k}\right|^{2}\cos^{2}\phi}N\frac{\partial k}{\partial\cos\phi}+\mathcal{O}\left(\frac{1}{N}\right).\end{equation}
 Introducing the density of states in the leads $\rho\left(\varepsilon\right)=\frac{1}{\pi}\frac{\partial k}{\partial\varepsilon},$
which for example, for a tight-binding lead with only nearest-neighbor
hopping $t_{0}$ and dispersion $\varepsilon_{k}=-2t_{0}\cos k$ equals
$1/\bigl(\pi\sqrt{4t_{0}^{2}-\varepsilon_{k}^{2}}\bigr)$, we finally
obtain \begin{equation}
\frac{\partial\arccos\left(\mp\left|t\left(\varepsilon_{k}\right)\right|\cos\phi\right)}{\partial\cos\phi}=\pi N\rho\left(\varepsilon_{k}\right)\frac{\partial\varepsilon_{k}}{\partial\cos\phi},\label{main}\end{equation}
where $t\left(\varepsilon_{k}\right)=t_{k}$. The condition Eq.~(\ref{cond1})
of validity can also be expressed in a form involving energy as a
variable\begin{equation}
N\gg\frac{1}{\rho\left(\varepsilon\right)}\left|\frac{\partial t\left(\varepsilon\right)}{\partial\varepsilon}\right|.\label{cond}\end{equation}
 Eq.~(\ref{main}) is the central result of this work. It expresses
the transmission probability $\left|t\left(\varepsilon\right)\right|^{2}$
of a sample connected to two leads in terms of the variation of single-electron
energy levels with magnetic flux penetrating the auxiliary ring system.
Employing the Landauer-Büttiker formula Eq.~(\ref{eq:landauer}),
this result also provides the zero-temperature conductance of the
system. From the derivation it is evident that the method becomes
exact as we approach the thermodynamic limit $N\to\infty$. 

\begin{figure}[htbp]
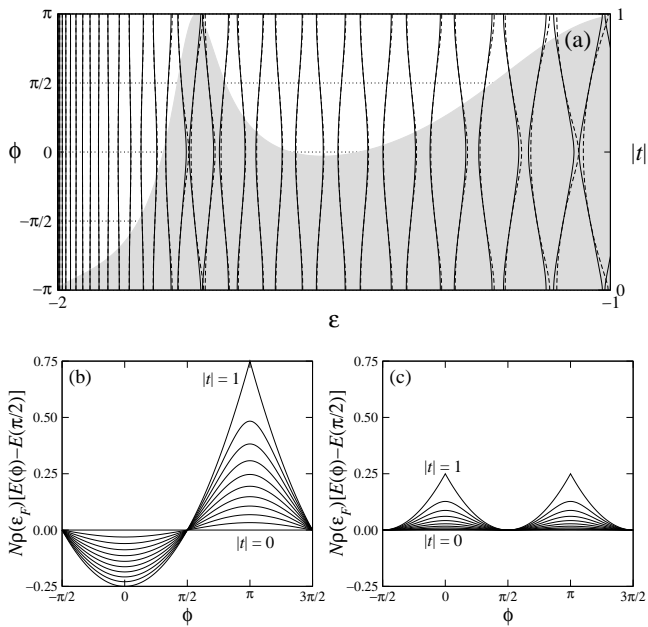

\begin{center}\includegraphics[%
  height=8.5cm,
  keepaspectratio,
  angle=-90,
  origin=lB]{Fig4a.eps}\end{center}

\begin{center}\includegraphics[%
  clip,
  width=8.5cm,
  keepaspectratio]{Fig4b.eps}\end{center}

\caption{\label{cap:eners}(a) Single-electron energy levels (full lines)
and the ground-state energies when a given single-electron level is
at the Fermi energy (dashed lines). Ground-state energies are shifted
so that both curves coincide for $\phi=\frac{\pi}{2}$. Note that
the energy curves are symmetric about $\phi=0$ as required by Eq.~(\ref{ereverse}).
The shaded area represents the magnitude of the transmission amplitude.
The system and the energy interval is the same as in Fig.~\ref{cap:figcos}.
(b, c) The large $N$ universal form of the ground-state energy vs.
flux curve for an even (b) and an odd (c) number of electrons in a
system. The magnitude of the transmission amplitude goes from 0 to
1 in steps of 0.1. }
\end{figure}

In general Eq.~(\ref{main}) has to be solved numerically to obtain
the transmission probability on a discrete set of energy points, one
for each energy level of a system. By increasing the system size $N$,
the density of these points increases and the errors decrease, as
the condition (\ref{cond}) is fulfilled better. We will return to
this point in Section~V where we consider the convergence issues
in detail. Here we present some special cases of Eq.~(\ref{main})
where analytic expressions can be obtained. By averaging the equation
over values of flux $\phi$ between $\phi=0$ and $\phi=\pi$ (note
that we may treat $\left|t\left(\varepsilon_{k}\right)\right|$ and
$\rho\left(\varepsilon_{k}\right)$ as constant while averaging as
the resulting error is of the order of $\frac{1}{N}$), we can relate
the transmission probability to the average magnitude of the derivative
of a single-electron energy with respect to the flux:\begin{equation}
\left|t\left(\varepsilon_{k}\right)\right|^{2}=\sin^{2}\left(\frac{\pi^{2}}{2}N\rho\left(\varepsilon_{k}\right)\overline{\left|\frac{\partial\varepsilon_{k}}{\partial\phi}\right|}\right).\label{average}\end{equation}
 Note that it is enough to calculate the energy levels at $\phi=0$
and $\phi=\pi$ to calculate the transmission probability as $\overline{\left|\partial\varepsilon_{k}/\partial\phi\right|}=\frac{1}{\pi}\left|\varepsilon_{k}\left(\pi\right)-\varepsilon_{k}\left(0\right)\right|$.
In Fig.~\ref{cap:eners}(a) it is illustrated how a large variation
of single-electron energy as the flux is changed from $\phi=0$ to
$\phi=\pi$ corresponds to a large conductance and vice versa. The
transmission probability can also be calculated from the derivative
at $\phi=\frac{\pi}{2}$ resulting in the second formula

\begin{equation}
\left|t\left(\varepsilon_{k}\right)\right|^{2}=\left(\pi N\rho\left(\varepsilon_{k}\right)\left.\frac{\partial\varepsilon_{k}}{\partial\phi}\right|_{\phi=\frac{\pi}{2}}\right)^{2}.\label{halfpi}\end{equation}
Again, Fig.~\ref{cap:eners}(a) shows that there is a correspondence
between a large sensitivity of a single-electron energy to the flux
at $\phi=\frac{\pi}{2}$, and a large conductance. Finally, we observe
that the curvature of energy levels at $\phi=0$ and $\phi=\pi$ also
gives information of conductance. The third formula reads\begin{equation}
\left|t\left(\varepsilon_{k}\right)\right|^{2}=1-\frac{1}{1+\left(\pi N\rho\left(\varepsilon_{k}\right)\left.\frac{\partial^{2}\varepsilon_{k}}{\partial\phi^{2}}\right|_{\phi=0,\pi}\right)^{2}}.\label{zeropi}\end{equation}

\subsection{Formulae relating conductance to the ground-state energy\label{sub:FormulaeGS}}

Above we showed how the flux variation of the energy of the last occupied
single-electron state allows one to calculate the zero-temperature
conductance through a noninteracting sample. The goal of this section
is to derive an alternative set of formulae, expressing the zero-temperature
conductance in terms of the flux variation of the ground-state energy
$E$, which for an even number of electrons in a noninteracting system
is simply a sum of single-electron energies up to the Fermi energy
$\varepsilon_{F}$, multiplied by 2 because of the electron's spin

\begin{equation}
E=2\sum_{\varepsilon_{n}\leq\varepsilon_{F}}\varepsilon_{n}.\label{energy}\end{equation}
We will show that the transmission probability at the Fermi energy
$\left|t\left(\varepsilon_{F}\right)\right|^{2}$ is related to the
ground-state energy of the ring system

\begin{equation}
\frac{1}{\pi}\frac{\partial\arccos^{2}\left(\mp\left|t\left(\varepsilon_{F}\right)\right|\cos\phi\right)}{\partial\cos\phi}=\pi N\rho\left(\varepsilon_{F}\right)\frac{\partial E}{\partial\cos\phi},\label{total}\end{equation}
 where the sign is $-$ and $+$ for an odd and an even number of
occupied single-electron states, respectively. The expression Eq.~(\ref{total})
is evidently correct if there are no electrons in the system, as it
gives a zero conductance in this case. To prove the formula for other
values of the Fermi energy, we use the principle of the mathematical
induction. To simplify the notation we introduce\begin{equation}
f_{s}\left(\varepsilon_{n}\right)=N\frac{\partial\varepsilon_{n}}{\partial\cos\phi}=\frac{1}{\pi\rho\left(\varepsilon_{n}\right)}\frac{\partial\arccos\left(s\left|t\left(\varepsilon_{n}\right)\right|\cos\phi\right)}{\partial\cos\phi},\end{equation}
\begin{equation}
F_{s}\left(\varepsilon_{n}\right)=N\frac{\partial E_{n}}{\partial\cos\phi}=\frac{1}{\pi^{2}\rho\left(\varepsilon_{n}\right)}\frac{\partial\arccos^{2}\left(s\left|t\left(\varepsilon_{n}\right)\right|\cos\phi\right)}{\partial\cos\phi},\end{equation}
 where $E_{n}$ is the ground-state energy of a system with the Fermi
energy at $\varepsilon_{n}$ and $s$ is either $1$ or $-1$, depending
on the signs in Eqs.~(\ref{main}) and (\ref{total}). Differentiating
the relation $E_{n}=E_{n-1}+2\varepsilon_{n}$ with respect to $\cos\phi$,
expressing the result in terms of functions $f_{s}$ and $F_{s}$
introduced above, and making use of the fact that the sign $s$ alternates
with $n$, we obtain \begin{equation}
F_{s}\left(\varepsilon_{n}\right)=F_{-s}\left(\varepsilon_{n-1}\right)+2f_{s}\left(\varepsilon_{n}\right).\end{equation}
 If we manage to show that this really is an identity, we have a proof
of Eq.~(\ref{total}). Using the exact relation $F_{s}\left(\varepsilon\right)-F_{-s}\left(\varepsilon\right)=2f_{s}\left(\varepsilon\right)$,
the expression transforms into \begin{equation}
F_{s}\left(\varepsilon_{n}\right)=F_{s}\left(\varepsilon_{n}\right)-\left[F_{-s}\left(\varepsilon_{n}\right)-F_{-s}\left(\varepsilon_{n-1}\right)\right].\end{equation}
 For a large number of sites $N$ in the ring and correspondingly,
a small separation of single-electron energy levels which is of the
order of $\frac{1}{N}$, the term in parenthesis equals $F_{-s}^{\prime}\left(\varepsilon_{n}\right)\left(\varepsilon_{n}-\varepsilon_{n-1}\right)$.
$F_{-s}^{\prime}\left(\varepsilon_{n}\right)$ can be factored into
$s\tilde{F}\left(\varepsilon_{n}\right)$ where $\tilde{F}\left(\varepsilon_{n}\right)$
does not depend on sign $s$. Therefore, although the term in parenthesis
in of the order of $\frac{1}{N}$, its sign alternates for successive
energy levels while its amplitude stays the same. Thus the error induced
by this term does not accumulate, it just adds an additional error
of the order of $\frac{1}{N}$ to the final result. 

In Fig.~\ref{cap:eners}(a), the variation of the ground-state energies
with magnetic flux is compared to the variation of the corresponding
single-electron energies. Note that as a consequence of Eq.~(\ref{total}),
the ground-state energy in the large $N$ limit takes a universal
form (see Fig.~\ref{cap:eners}(b))

\begin{eqnarray}
 &  & \!\!\!\!\!\! E\left(\phi\right)-E\left(\frac{\pi}{2}\right)=\nonumber \\
 &  & \!\!\!\!\!\!\quad=\frac{1}{\pi^{2}N\rho\left(\varepsilon_{F}\right)}\Bigl(\arccos^{2}\left(\mp\left|t\left(\varepsilon_{F}\right)\right|\cos\phi\right)-\frac{\pi^{2}}{4}\Bigr).\quad\quad\label{eq:even}\end{eqnarray}
For systems with an odd number of electrons, the ground-state energy
is obtained by adding a single-electron energy corresponding to Eq.~(\ref{main})
and the universal form reads (see Fig.~\ref{cap:eners}(c))

\begin{equation}
E\left(\phi\right)-E\left(\frac{\pi}{2}\right)=\frac{1}{\pi^{2}N\rho\left(\varepsilon_{F}\right)}\arcsin^{2}\left(\left|t\left(\varepsilon_{F}\right)\right|\cos\phi\right).\label{eq:odd}\end{equation}

In general, Eq.~(\ref{total}) can only be solved numerically to
obtain the transmission probability. However, as was the case for
single-electron energies, analytic solutions can be found in certain
special cases. The derivative of the ground-state energy with respect
to flux gives the persistent current in the ring $j=\frac{e}{\hbar}\frac{\partial E}{\partial\phi}$
\cite{Bloch65,Buttiker83}. Using the Landauer-Büttiker formula Eq.~(\ref{eq:landauer}),
one can calculate the conductance from the flux averaged magnitude
of the persistent current in the system\begin{equation}
\left|t\left(\varepsilon_{F}\right)\right|^{2}=\sin^{2}\left(\frac{\pi^{2}\hbar}{2e}N\rho\left(\varepsilon_{F}\right)\overline{\left|j\left(\phi\right)\right|}\right).\label{eq:averageS}\end{equation}
Only two ground-state energy calculations need to be performed to
obtain the conductance as $\frac{\hbar}{e}\overline{\left|j\left(\phi\right)\right|}=\frac{1}{\pi}\left|E\left(\pi\right)-E\left(0\right)\right|$.
This formula was also discussed in Refs.~\cite{Favand98,Molina02}
for the case where the transmission probability is small. The second
formula relates the conductance to the persistent current at $\phi=\frac{\pi}{2}$\cite{Sushkov01,Meden02},\begin{equation}
\left|t\left(\varepsilon_{F}\right)\right|^{2}=\left(\frac{\pi\hbar}{e}N\rho\left(\varepsilon_{F}\right)j\left(\frac{\pi}{2}\right)\right)^{2}.\label{eq:halfpiS}\end{equation}
The third formula, corresponding to Eq.~(\ref{zeropi}) in the single-electron
case, turns out to be more complicated and gives an implicit relation
for $\left|t\left(\varepsilon_{F}\right)\right|$\begin{eqnarray}
 &  & \pi N\rho\left(\varepsilon_{F}\right)\left.\frac{\partial^{2}E}{\partial\phi^{2}}\right|_{\mathrm{min,}\,\mathrm{max}}=\nonumber \\
 &  & \quad=\pm\frac{2\left|t\left(\varepsilon_{F}\right)\right|}{\pi\sqrt{1-\left|t\left(\varepsilon_{F}\right)\right|^{2}}}\arccos\left(\pm\left|t\left(\varepsilon_{F}\right)\right|\right).\label{eq:drude}\end{eqnarray}
 Here the upper and the lower signs correspond to the second derivative
at a minimum and at a maximum of the energy vs. flux curve, respectively.
Minima (maxima) occur at $\phi=0\left(\pi\right)$ if an odd number
of single-electron levels is occupied and at $\phi=\pi\left(0\right)$
if an even number of levels is occupied. The second derivative in
a minimum is proportional to the charge stiffness $D=\frac{N}{2}\left.\partial^{2}E/\partial\phi^{2}\right|_{\mathrm{min}}$
of the system\cite{Kohn64,Prelovsek01}. We can also define the corresponding
quantity for a maximum as $\tilde{D}=-\frac{N}{2}\left.\partial^{2}E/\partial\phi^{2}\right|_{\mathrm{max}}$.
In general, Eq.~(\ref{eq:drude}) has to be solved numerically. However,
in the limit of a very small conductance and in the vicinity of the
unitary limit, additional analytic formulae are valid\begin{equation}
\left|t\left(\varepsilon_{F}\right)\right|=\left\{ \begin{array}{ll}
2\pi\rho\left(\varepsilon_{F}\right)D, & \left|t\left(\varepsilon_{F}\right)\right|\rightarrow0,\\
\frac{1}{2}+\frac{3\pi}{4}2\pi\rho\left(\varepsilon_{F}\right)D, & \left|t\left(\varepsilon_{F}\right)\right|\rightarrow1.\end{array}\right.\end{equation}
 Note that there is a quadratic relation between the conductance and
the charge stiffness in the low conductance limit. The corresponding
formulae for the maximum of the energy vs. flux curve are\begin{equation}
\left|t\left(\varepsilon_{F}\right)\right|=\left\{ \begin{array}{ll}
2\pi\rho\left(\varepsilon_{F}\right)\tilde{D}, & \left|t\left(\varepsilon_{F}\right)\right|\rightarrow0,\\
1-\frac{2}{\left(2\pi\rho\left(\varepsilon_{F}\right)\tilde{D}\right)^{2}}, & \left|t\left(\varepsilon_{F}\right)\right|\rightarrow1.\end{array}\right.\end{equation}
A detailed analysis of convergence properties of the formulae derived
in this Section is presented Section V.

We stress again that the validity of these formulae is based on an
assumption that the number of sites in the ring is sufficiently large
according to the condition Eq.~(\ref{cond}). This means that if
$t(\varepsilon)$ exhibits sharp resonances, the calculation has to
be performed on such a large auxiliary ring system that in the energy
interval of interest (the width of the resonance) there is a large
number of eigenenergies $\varepsilon_{n}$. Then $t(\varepsilon)\sim t(\varepsilon_{n'})$,
where $\varepsilon_{n'}$ is the eigenenergy closest to $\varepsilon$.
Such sharp resonances in $t(\varepsilon)$ are expected e.g. in chaotic
systems~\cite{Jalabert92,Verges98}. The present method might be
impractical (but still correct) in this case.

\section{Conductance of an interacting system}

The zero-temperature conductance of a noninteracting system can thus
be determined with the transmission probability obtained from one
of the formulae we derived in the previous Section, and the Landauer-Büttiker
formula. The main challenge, however, remains the question of the
validity of this type of approach for interacting systems. In this
Section we give arguments why the approach is correct for a class
of interacting systems exhibiting Fermi liquid properties. In order
to reach this goal, we present four essential steps as follows.

\subsection*{Step 1: Conductance of a Fermi liquid system at $\bm T\bm=\bm0$}

The basic property that characterizes Fermi liquid systems\cite{Nozieres64}
is that the states of a noninteracting system of electrons are continuously
transformed into states of the interacting system as the interaction
strength increases from zero to its actual value. One can then study
the properties of such a system by means of the perturbation theory,
regarding the interaction strength as the perturbation parameter.
The concept of the Fermi liquid was first introduced for translation-invariant
systems by Landau\cite{Landau56,Landau57}, and was later also extended
to systems of the type we study here\cite{Nozieres74}. 

The linear response conductance of a general interacting system of
the type shown in Fig.~\ref{cap:System} can be calculated from the
Kubo formula~\cite{Bruus02,Oguri01}\begin{equation}
G=\lim_{\omega\rightarrow0}\frac{ie^{2}}{\omega+i\delta}\Pi_{II}\left(\omega+i\delta\right),\end{equation}
where $\Pi_{II}\left(\omega+i\delta\right)$ is the retarded current-current
correlation function\begin{equation}
\Pi_{II}\left(t-t^{\prime}\right)=-i\theta\left(t-t^{\prime}\right)\left\langle \left[I\left(t\right),I\left(t^{\prime}\right)\right]\right\rangle .\end{equation}
For Fermi liquid systems at $T=0$, the current-current correlation
function can be expressed in terms of the Green's function $G_{n^{\prime}n}\left(z\right)$
of the system and the conductance is given with~\cite{Oguri01} 

\begin{equation}
G=\frac{2e^{2}}{h}\left|\frac{1}{-i\pi\rho\left(\varepsilon_{F}\right)}e^{-ik_{F}\left(n^{\prime}-n'\right)}G_{n^{\prime}n}\left(\varepsilon_{F}+i\delta\right)\right|^{2},\label{eq:GFisherLee}\end{equation}
where $n$ and $n^{\prime}$ are sites in the left and the right lead,
respectively. One can \emph{define} the transmission amplitude as\begin{equation}
t\left(\varepsilon\right)\equiv\frac{1}{-i\pi\rho\left(\varepsilon\right)}e^{-ik\left(n^{\prime}-n\right)}G_{n^{\prime}n}\left(\varepsilon+i\delta\right),\label{eq:FermiLB2}\end{equation}
and the conductance formula Eq.~(\ref{eq:GFisherLee}) then reads\begin{equation}
G=\frac{2e^{2}}{h}\left|t\left(\varepsilon_{F}\right)\right|^{2}.\label{eq:FermiLB1}\end{equation}
For non-interacting systems, $t\left(\varepsilon\right)$ defined
this way reduces to the standard transmission amplitude (Fisher-Lee
relation~\cite{Fisher81}) and Eq.~(\ref{eq:FermiLB1}) represents
the Landauer-Büttiker formula. In the next step, we will show that
the transmission amplitude Eq.~(\ref{eq:FermiLB2}) has a direct
physical interpretation also for interacting systems, being the transmission
amplitude of Fermi liquid quasiparticles.

\subsection*{Step 2: Quasiparticle Hamiltonian }

In this step, we generalize the quasiparticle approximation to the
Green's function, presented for the single-impurity Anderson model
in Ref.~\cite{Hewson93}, to the case where the interaction is present
in more than a single site.

In Fermi liquid systems obeying the time-reversal symmetry, the imaginary
part of the retarded self-energy at $T=0$ vanishes at the Fermi energy
and is quadratic for frequencies close to the Fermi energy\cite{Yamada86,Oguri97}.
Using the Fermi energy as the origin of the energy scale, i.e. $\omega-\varepsilon_{F}\rightarrow\omega$,
we can express this as \begin{equation}
\mathrm{Im}\bm\Sigma\left(\omega+i\delta\right)\propto\omega^{2}.\label{Luttinger}\end{equation}
Close to the Fermi energy, the self-energy can be expanded in powers
of $\omega$ resulting in an approximation to the Green's function,\begin{eqnarray}
 &  & \mathbf{G}^{-1}\left(\omega+i\delta\right)=\omega\mathbf{1}-\mathbf{H}^{\left(0\right)}-\bm\Sigma\left(0+i\delta\right)-\label{eq:expansion}\\
 &  & \quad-\omega\left.\frac{\partial\bm\Sigma\left(\omega+i\delta\right)}{\partial\omega}\right|_{\omega=0}+\mathcal{O}\left(\omega^{2}\right).\end{eqnarray}
 Here $\mathbf{H}^{\left(0\right)}$ contains matrix elements of the
noninteracting part of the Hamiltonian (\ref{eq:H0plusU}). Note that
expansion coefficients are real because of Eq.~(\ref{Luttinger}).
Let us introduce the renormalization factor matrix $\mathbf{Z}$ as\begin{equation}
\mathbf{Z}^{-1}=\mathbf{1}-\left.\frac{\partial\bm\Sigma\left(\omega+i\delta\right)}{\partial\omega}\right|_{\omega=0}.\label{eq:zzzz}\end{equation}
 The Green's function for $\omega$ close to the Fermi energy can
then be expressed as\begin{equation}
\mathbf{G}^{-1}\left(\omega+i\delta\right)=\mathbf{Z}^{-1/2}\tilde{\mathbf{G}}^{-1}\left(\omega+i\delta\right)\mathbf{Z}^{-1/2}+\mathcal{O}\left(\omega^{2}\right)\!,\!\!\label{eq:gfermi}\end{equation}
where we defined the quasiparticle Green's function\begin{equation}
\tilde{\mathbf{G}}^{-1}\left(\omega+i\delta\right)=\omega\mathbf{1}-\tilde{\mathbf{H}}\end{equation}
as the Green's function of a \emph{noninteracting} quasiparticle Hamiltonian

\begin{equation}
\tilde{\mathbf{H}}=\mathbf{Z}^{1/2}\left[\mathbf{H}^{\left(0\right)}+\bm\Sigma\left(0+i\delta\right)\right]\mathbf{Z}^{1/2}.\label{quasiham}\end{equation}
 Note that factoring the renormalization factor matrix as we did above
ensures the hermiticity of the resulting quasiparticle Hamiltonian. 

Matrix elements of $\mathbf{Z}$ differ from those of an identity
matrix only if they correspond to sites of the central region. In
other cases, as the interaction is limited to the central region,
the corresponding self-energy matrix element is zero. Therefore, comparing
the quasiparticle Hamiltonian to the noninteracting part of the real
Hamiltonian, we observe that the effect of the interaction is to renormalize
the matrix elements of the central region Hamiltonian (\ref{eq:HCni})
and those corresponding to the hopping between the central region
and the leads (\ref{eq:Hhop}). The values of the renormalized matrix
elements depend on the value of the Fermi energy of the system.

\begin{figure}[htbp]
\begin{center}\includegraphics[%
  width=6cm,
  keepaspectratio]{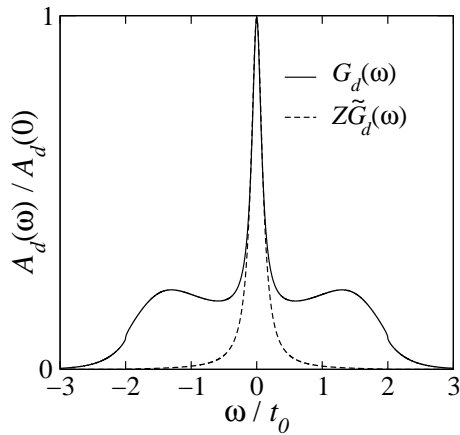}\end{center}

\caption{\label{cap:sigma}The $T=0$ local spectral function and the corresponding
quasiparticle approximation for the Anderson impurity model as shown
in Fig.~\ref{cap:andsys}. The values of parameters are $t_{1}=0.4t_{0}$,
$U=1.92t_{0}$ and $\varepsilon_{d}=-\frac{U}{2}$. The calculations
were performed within the second-order perturbation theory as described
in Appendix A. }
\end{figure}

Let us illustrate the ideas introduced above for the case of the standard
Anderson impurity model~\cite{Hewson93}. We calculated the self-energy
in the second-order perturbation theory approximation~\cite{Horvatic80,Horvatic85,Horvatic87}
and constructed the quasiparticle Hamiltonian according to Eq.~(\ref{quasiham}).
In Fig.~\ref{cap:sigma} the local spectral functions corresponding
to both the original interacting Hamiltonian and the noninteracting
quasiparticle Hamiltonian are presented. The agreement of both results
is perfect in the vicinity of the Fermi energy where the expansion
(\ref{eq:expansion}) is valid. 

The reason for introducing the quasiparticle Hamiltonian is to obtain
an alternative expression for the conductance in terms of the quasiparticle
Green's function. Eq.~(\ref{eq:gfermi}) relates the values of the
true and the quasiparticle Green's function at the Fermi energy, \begin{equation}
\mathbf{G}\left(0+i\delta\right)=\mathbf{Z}^{1/2}\tilde{\mathbf{G}}\left(0+i\delta\right)\mathbf{Z}^{1/2}.\label{eq:zgtildez}\end{equation}
Specifically, if both $n$ and $n^{\prime}$ are sites in the leads,
$G_{n^{\prime}n}\left(0+i\delta\right)=\tilde{G}_{n^{\prime}n}\left(0+i\delta\right)$
as a consequence of the properties of the renormalization factor matrix
$\mathbf{Z}$ discussed above. Eq.~(\ref{eq:FermiLB2}) then tells
us that the zero-temperature conductance of a Fermi liquid system
is identical to the zero-temperature conductance of a noninteracting
system defined with the quasiparticle Hamiltonian for a given value
of the Fermi energy.

\subsection*{Step 3: Quasiparticles in a finite system}

The conclusions reached in the first two steps are based on an assumption
of the thermodynamic limit, i.e. they are valid if the central region
is coupled to semiinfinite leads. Here we generalize the concept of
quasiparticles to a finite ring system with $N$ sites and $M$ electrons,
threaded by a magnetic flux $\phi$. Let us define the quasiparticle
Hamiltonian for such a system,\begin{equation}
\tilde{\mathbf{H}}\left(N,\phi;M\right)=\mathbf{Z}^{1/2}\left[\mathbf{H}^{\left(0\right)}\left(N,\phi\right)+\bm\Sigma\left(0+i\delta\right)\right]\mathbf{Z}^{1/2}.\label{eq:assumptquasi}\end{equation}
 Here the self-energy and the renormalization factor matrix are determined
in the thermodynamic limit where, as we prove in Appendix A, they
are independent of $\phi$ and correspond to those of an infinite
two-lead system. 

Suppose now that we knew the exact values of the renormalized matrix
elements in the quasiparticle Hamiltonian (\ref{eq:assumptquasi}).
As this is a noninteracting Hamiltonian, we could then apply the conductance
formulae of the previous Section to calculate the zero-temperature
conductance of an infinite two-lead system with the same central region
and central region-lead hopping matrix elements, i.e. of a system
described with the quasiparticle Hamiltonian (\ref{quasiham}). As
shown in step 2, this procedure would provide us with the exact conductance
of the original interacting system. However, to obtain the values
of the renormalized matrix elements, one needs to calculate the self-energy
of the system, which is a difficult many-body problem. In the next
step, we will show, that there is an alternative and easier way to
achieve the same goal. 

\begin{figure}[htbp]
\begin{center}\includegraphics[%
  width=8.5cm,
  keepaspectratio]{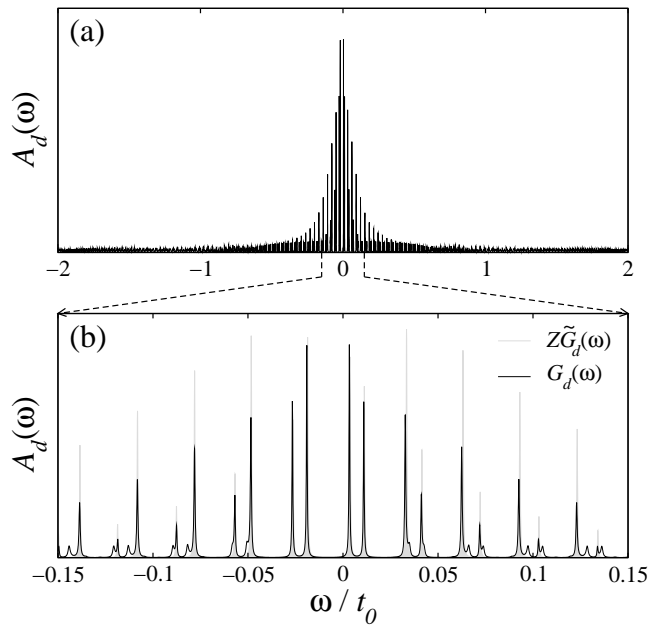}\end{center}

\caption{\label{cap:sigma1}(a) The $T=0$ local spectral function as in Fig.~\ref{cap:sigma},
but for a ring system with $N=400$ sites and flux $\phi=\frac{3\pi}{4}$.
(b) The spectral function in the vicinity of the Fermi energy (dashed
lines) compared to that corresponding to the quasiparticle Hamiltonian
(\ref{eq:assumptquasi}). Both the spectral density of the interacting
system and the matrix elements of the quasiparticle Hamiltonian were
calculated within the second order perturbation theory.}
\end{figure}

In Fig.~\ref{cap:sigma1} we compare the spectral density of an Anderson
impurity embedded in a finite ring system to that of the corresponding
quasiparticle Hamiltonian (\ref{eq:assumptquasi}). Note that the
spectral density of the quasiparticle Hamiltonian correctly describes
the true spectral density in the vicinity of the Fermi energy.

\subsection*{Step 4: Validity of the conductance formulae\label{sub:Step4}}

In this last step we finally show how to calculate the conductance
of an interacting system. In Appendix A we study the excitation spectrum
of a finite ring system threaded with a magnetic flux and containing
a region with interaction. We show that \begin{eqnarray}
 &  & E\left[N,\phi;M+1\right]-E\left[N,\phi;M\right]=\nonumber \\
 &  & \quad=\tilde{\varepsilon}\left(N,\phi;M;1\right)+\mathcal{O}\left(N^{-\frac{3}{2}}\right),\label{eq:assumpt}\end{eqnarray}
where $E\left(N,\phi;M\right)$ and $E\left(N,\phi;M+1\right)$ are
the ground-state energies of the interacting Hamiltonian for a ring
system with $N$ sites and flux $\phi$, containing $M$ and $M+1$
electrons, respectively, and $\tilde{\varepsilon}\left(N,\phi;M;1\right)$
is the energy of the first single-electron level above the Fermi energy
of the finite ring quasiparticle Hamiltonian (\ref{eq:assumptquasi}).
This estimation allows one to use single-electron formulae of Sec.~\ref{sub:FormulaeSE}
to calculate the zero-temperature conductance for a Fermi liquid system.
We showed in step 3 that inserting $\tilde{\varepsilon}\left(N,\phi;M;1\right)$
into these formulae would give us the correct conductance. Eq.~(\ref{eq:assumpt})
proves, that the same result is obtained if the difference of the
ground state energies of an interacting system $E\left[N,\phi;M+1\right]-E\left[N,\phi;M\right]$
is inserted into a formula instead. The estimated error, which is
of the order of $N^{-\frac{3}{2}}$, is for a large $N$ negligible,
because it is much smaller than the quasiparticle level spacing, which
is of the order of $\frac{1}{N}$.

As demonstrated in Sec.~\ref{sub:FormulaeGS}, the conductance of
a noninteracting system can also be calculated from the variation
of the ground-state energy with flux through the ring. The proof of
the formulae involved only the properties of a set of neighboring
single-electron energy levels. We assumed the validity of single-electron
conductance formulae for each of these levels and made use of the
fact that the ground-state energy of the system increases by a sum
of the relevant single-electron energies as the levels become occupied
with additional electrons. For Fermi liquid systems, the first assumption
was proved above. The second assumption, which for noninteracting
systems is obvious, is proved in Appendix A. There we show that as
a finite number of additional electrons is added to an interacting
system, the successive ground-state energies are determined by the
single-electron energy levels of the same quasiparticle Hamiltonian
with a very good accuracy (\ref{eq:energyassumpt}). Therefore, the
proof of Sec.~\ref{sub:FormulaeGS} is also valid for interacting
Fermi liquid systems, provided a system is the Fermi liquid for all
values of the Fermi energy below its actual value.

\section{Numerical tests of the method}

\subsection{Noninteracting system}

\begin{figure}[htbp]
\begin{center}\includegraphics[%
  width=6.7cm,
  keepaspectratio]{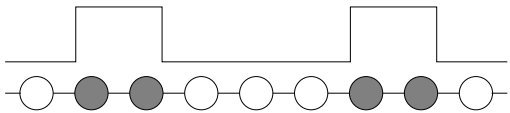}\end{center}

\caption{\label{cap:testnisys}A double barrier noninteracting system. The
height of the barriers is $0.5t_{0}$, where $t_{0}$ is the hopping
matrix element between neighboring sites.}
\end{figure}

\begin{figure*}[htbp]
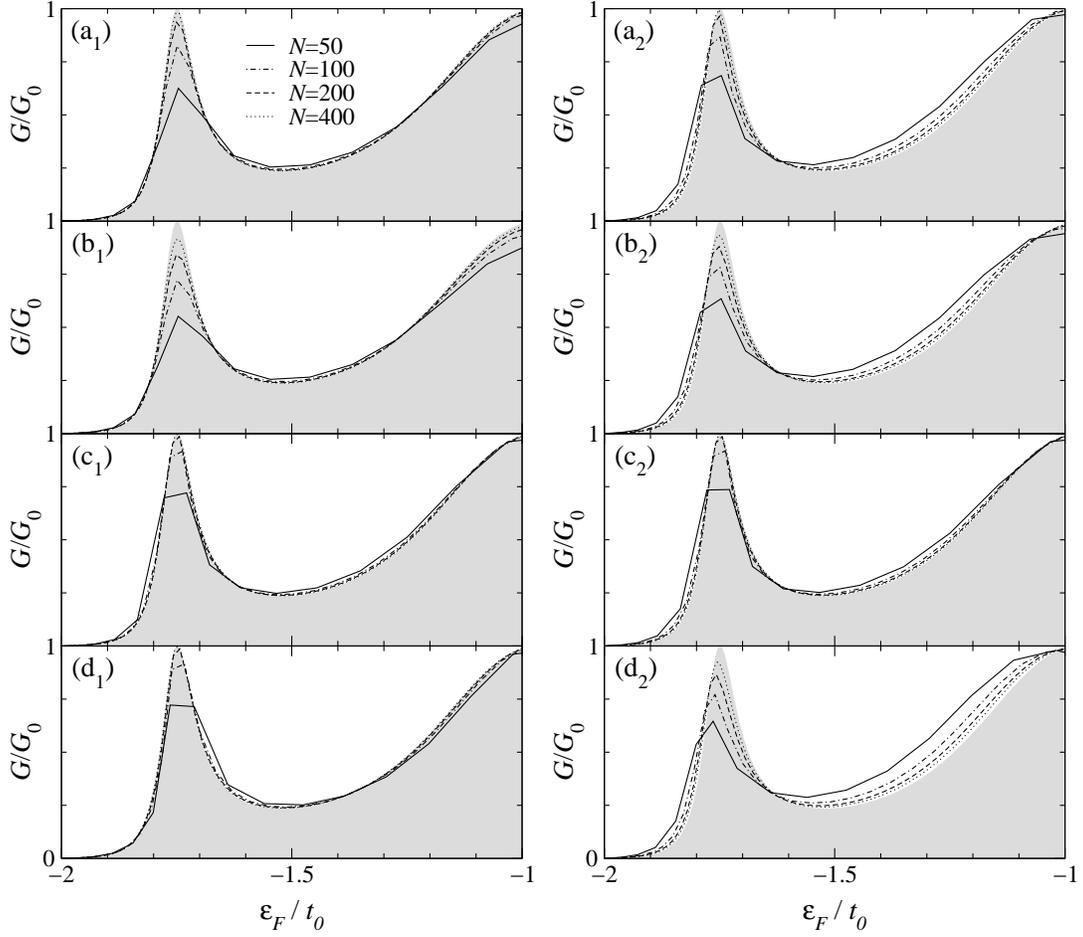

\begin{center}\includegraphics[%
  width=7cm,
  keepaspectratio]{Fig8a.eps}~~\includegraphics[%
  width=7cm,
  keepaspectratio]{Fig8b.eps}\end{center}

\caption{\label{cap:testni}Exact and approximate zero-temperature conductance
vs. Fermi energy curves for the system in Fig.~\ref{cap:testnisys}.
The shaded area shows the exact result. The left set of figures shows
the approximations obtained using the single-electron formulae of
Sec.~\ref{sub:FormulaeSE} while the right set of figures corresponds
to the ground-state energy formulae of Sec.~\ref{sub:FormulaeGS}.
Different curves correspond to different number of sites $N$ in the
ring. In figures ($\mathrm{a}_{1}$) and ($\mathrm{a}_{2}$) the conductance
was calculated using Eqs.~(\ref{average}) and (\ref{eq:averageS}),
in figures ($\mathrm{b}_{1}$) and ($\mathrm{b}_{2}$) using Eqs.~(\ref{halfpi})
and (\ref{eq:halfpiS}), while in the other figures Eqs.~(\ref{zeropi})
and (\ref{eq:drude}) were used, in ($\mathrm{c}_{1}$) and ($\mathrm{c}_{2}$)
applied to the maximum and in ($\mathrm{d}_{1}$) and ($\mathrm{d}_{2}$)
to the minimum of energy vs. flux curves.}
\end{figure*}

In this Section we discuss the convergence properties of the conductance
formulae derived in Sec.~\ref{sec:sec23}. As a test system we use
a double-barrier potential scattering problem presented in Fig.~\ref{cap:testnisys}.
Results of various formulae for different number of sites in the ring
are presented in Fig.~\ref{cap:testni}. The exact zero-temperature
conductance for this system exhibits a sharp resonance peak superimposed
on a smooth background conductance. We notice immediately that as
the number of sites in the ring increases, the convergence is generally
faster in the region where the conductance is smooth than in the resonance
region, which is consistent with the condition (\ref{cond}). Comparing
the results obtained employing different conductance formulae we observe
that the convergence is the fastest in both the single-electron and
the ground-state energy case if the formulae of Eqs.~(\ref{zeropi})
and (\ref{eq:drude}) are applied to the maximum of the energy vs.
flux curve (or to the minimum in the single-electron case). Formulae
of Eqs.~(\ref{average}) and (\ref{eq:averageS}) expressing the
conductance in terms of the difference of the energies at $\phi=0$
and $\phi=\pi$ converge somewhat slower. Note however that in the
former case the second derivative of the energy with respect to the
flux has to be evaluated while in the later, the energy difference
is large and because of that, the calculation is much more well behaved.
From the computational point of view there is another advantage of
the energy difference formulae. In this case, all the matrix elements
can be made real if one chooses such a vector potential that only
one hopping matrix element if modified by the flux as then the additional
phase factor is $e^{\pm i\pi}=-1$. Finally, the remaining formulae,
employing the slope of the energy vs. flux curve at $\phi=\frac{\pi}{2}$
and the curvature in the minimum of the ground-state energy vs. flux
curve, do not show convergence properties comparable to those of the
formulae discussed above.

\subsection{Anderson impurity model}

\begin{figure}[htbp]
\begin{center}\includegraphics[%
  width=5.2cm,
  keepaspectratio]{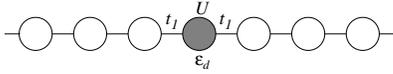}\end{center}

\caption{\label{cap:andsys}The Anderson impurity model realized as a quantum
dot coupled to two leads. The dot is described with the energy level
$\varepsilon_{d}$ and the Coulomb energy of a doubly occupied level
$U$. $t_{1}$ is the hopping between the dot and leads.}
\end{figure}

In 1980s several theories~\cite{Glazman88,Ng88} were put forward
proposing a realization of the Anderson impurity model~\cite{Anderson61}
in systems consisting of a quantum dot coupled to two leads (see Fig.~\ref{cap:andsys}).
These theories show that the topmost occupied energy level in a quantum
dot with an odd number of electrons can be associated with the Anderson
model $\varepsilon_{d}$ level and such a system should mimic the
old Kondo problem of a magnetic spin $\frac{1}{2}$ impurity in a
metal host. In recent years signatures of the Kondo physics in electron
transport through quantum dots have also been found experimentally~\cite{Goldhaber98,Wiel00}.
The Anderson model is well defined and is an attractive testing ground
for new numerical and analytical methods that are developed to tackle
other challenging many-body problems. Therefore, we will also take
it as a nontrivial example to test results of the conductance formulae
we derived in this paper.

There are three distinct parameter regimes of the Anderson model.
If $\varepsilon_{d}<\varepsilon_{F}<\varepsilon_{d}+U$ with $\left|\varepsilon_{d}+U-\varepsilon_{F}\right|\gg\Delta$
and $\left|\varepsilon_{d}-\varepsilon_{F}\right|\gg\Delta$, where
$\Delta$ is the coupling of the quantum dot to leads, we are in the
Kondo regime. In this regime, a narrow Kondo resonance is formed in
the spectral function at the Fermi energy for temperatures below and
close to the Kondo temperature, which corresponds to the width of
the resonance. The zero-temperature conductance in the Kondo regime
reaches the unitary limit of $\frac{2e^{2}}{h}$. Letting either $\varepsilon_{d}$
or $\varepsilon_{d}+U$ approach the Fermi energy so that $\left|\varepsilon_{d}+U-\varepsilon_{F}\right|$
or $\left|\varepsilon_{d}-\varepsilon_{F}\right|$ becomes comparable
with $\Delta$, we enter the mixed valence regime where the charge
fluctuations on the dot become important. In this regime, the resonance
becomes wider and merges with the resonance corresponding to $\varepsilon_{d}$
or $\varepsilon_{d}+U$ levels. More important for our discussion
is the fact that the resonance moves away from the Fermi energy and
therefore, the conductance drops as we enter this regime. Finally,
there are two nonmagnetic regimes, one in which the {}``impurity''
level is predominately empty, $\varepsilon_{d}-\varepsilon_{F}\gg\Delta$,
known as the empty orbital regime, and the corresponding regime where
the dot is doubly occupied. In these regimes, the conductance drops
toward zero.

\begin{figure}[htbp]
\begin{center}\includegraphics[%
  width=8.5cm,
  keepaspectratio]{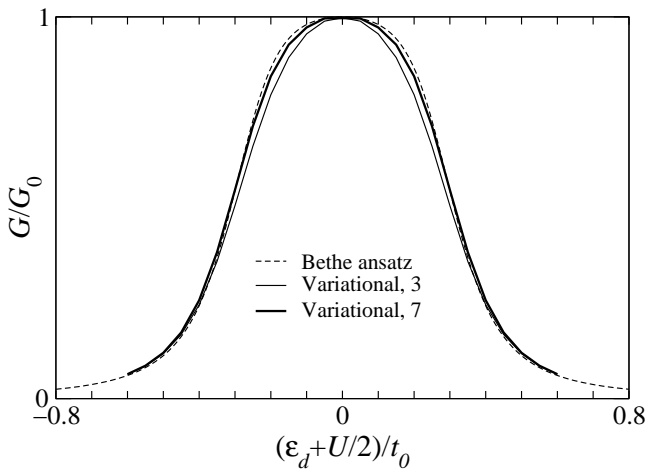}\end{center}

\caption{\label{cap:andvar}The zero-temperature conductance calculated from
ground-state energy vs. magnetic flux in a finite ring system using
the variational method of Appendix B with 3 and 7 basis functions.
For comparison, the exact Bethe ansatz result is presented with a
dashed line. The system shown in Fig.~\ref{cap:andsys} was used,
with $U=0.64t_{0}$ and $t_{1}=0.2t_{0}$.}
\end{figure}

\begin{figure}[htbp]
\begin{center}\includegraphics[%
  width=8.5cm,
  keepaspectratio]{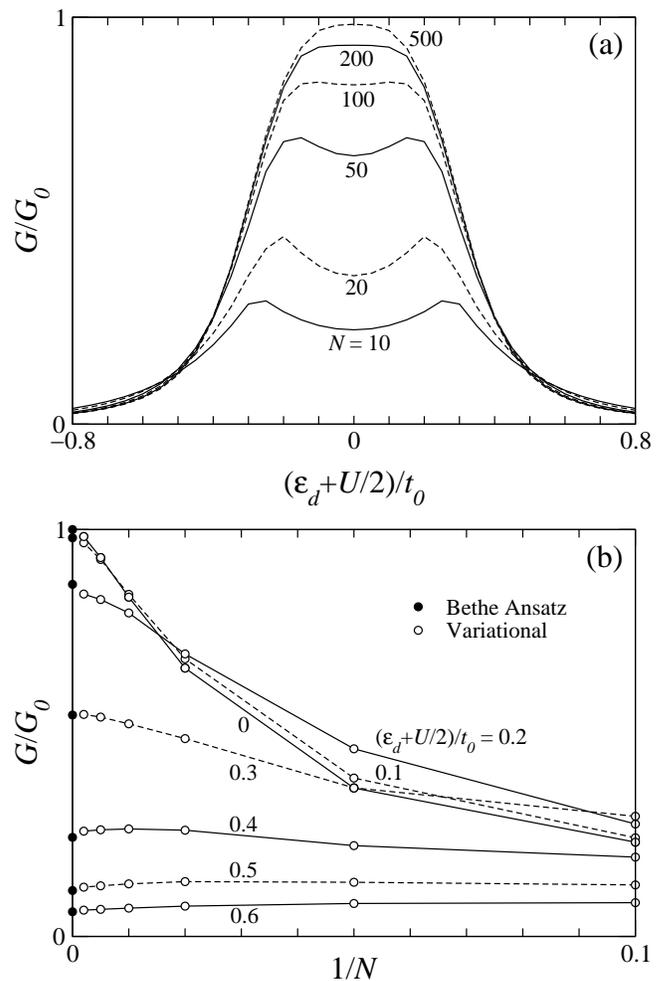}\end{center}

\caption{\label{cap:andfss} (a) Results of conductance calculations using
Eq.~(\ref{eq:averageS}) for the system presented in Fig.~\ref{cap:andvar}
as the number of sites in the ring increases. Note that the convergence
is the fastest in the empty orbital regime and the slowest in the
Kondo regime. (b) Finite-size scaling analysis of the same results
for various values of $\varepsilon_{d}$. With black dots, the Bethe
ansatz values are shown. Energies were calculated using the variational
method of Appendix B with 7 variational basis functions.}
\end{figure}

In Fig.~\ref{cap:andvar} the zero-temperature conductance through
a quantum dot acting as an Anderson impurity is presented and compared
to exact results of the Bethe ansatz approach\cite{Wiegman82a,Wiegman83}.
To calculate the conductance, Eq.~(\ref{eq:averageS}) was used,
with the ground-state energies at $\phi=0$ and $\phi=\pi$ obtained
using the variational method presented in Appendix B. There are two
variational parameters defining the auxiliary Hamiltonian (\ref{auxillary}),
one describing the effective energy level on the dot and the other
renormalizing hoppings into the leads. Two different variational basis
sets were used in calculations. In the first set, the basis consisted
of wavefunctions (\ref{basis}). As a result of the rotational symmetry
in the spin degree of freedom, two of the basis functions may be merged
into one. Therefore, the basis set consisted of projections of the
auxiliary Hamiltonian ground state $\left|\tilde{0}\right\rangle $
to states with empty $P_{0}\left|\tilde{0}\right\rangle $, singly
occupied $P_{1}\left|\tilde{0}\right\rangle =P_{\uparrow}\left|\tilde{0}\right\rangle +P_{\downarrow}\left|\tilde{0}\right\rangle $
and doubly occupied $P_{2}\left|\tilde{0}\right\rangle $ dot level.
In the second basis set, wavefunctions $P_{1}VP_{0}\left|\tilde{0}\right\rangle ,$
$P_{0}VP_{1}\left|\tilde{0}\right\rangle $, $P_{2}VP_{1}\left|\tilde{0}\right\rangle $
and $P_{1}VP_{2}\left|\tilde{0}\right\rangle $ (\ref{eq:extbas})
were added to those of the first set, with $V=V_{L}+V_{R}$ being
the operator describing the hopping between the dot and the leads
(\ref{eq:Hhop}). For each position of the $\varepsilon_{d}$ level
relative to the Fermi energy, we increased the number of sites in
the ring until the conductance converged. The number of sites needed
to achieve convergence (see Fig.~\ref{cap:andfss}a) was the lowest
in the empty orbital regime and the highest (about 1000 for the system
shown in Fig.~\ref{cap:andvar}) in the Kondo regime. This is a consequence
of Eq.~(\ref{cond}) as a narrow resonance related to the Kondo resonance
appears in the transmission probability of the quasiparticle Hamiltonian
(\ref{quasiham}) in the Kondo regime. In the mixed valence regime,
the width of the resonance becomes comparable to $\Delta$, which
is much larger than the Kondo temperature and the convergence is thus
faster. In the empty orbital regime the resonance moves away from
the Fermi energy and an even smaller number of sites is needed to
achieve convergence. 

Let us return to results shown in Fig.~\ref{cap:andvar}. Note that
extending the variational space from 3 to 7 basis functions significantly
improves the agreement with the exact result. The remaining discrepancy
at the larger basis set can be attributed to the approximate nature
of the variational method. Another source of error could be the fact
that the Bethe ansatz solution assumes there is a constant coupling
to an infinitely wide conduction band. In our case, the conduction
band is formed by the states in a tight-binding ring, the coupling
to which is not constant. However, it is almost constant in the energy
interval we are interested in, i.e. near the center of the band. In
order to estimate the effect of the nonconstant coupling on the conductance,
we calculated the dot occupation number within the second order perturbation
theory for both the case of a constant coupling and for the case of
a tight-binding ring. We then calculated the conductance in each case
making use of the Friedel sum rule~\cite{Langreth66}. The agreement
is significantly better than the difference between the Bethe ansatz
and variational conductance curves in Fig.~\ref{cap:andvar}. Therefore,
we believe that the use of Bethe ansatz results is justified for this
particular problem.

In Fig.~\ref{cap:andfss}b a finite-size scaling analysis of the
convergence is presented. Note that for rings with a large number
of sites $N$, the error scales approximately as $\frac{1}{N}$.

\subsection{Double quantum dot}

\begin{figure}[b]
\begin{center}\includegraphics[%
  width=6cm,
  keepaspectratio]{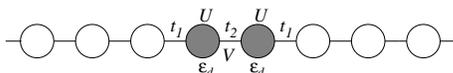}\end{center}

\caption{\label{cap:hubsys}A double quantum dot system. Each of the dots
with energy level $\varepsilon_{d}$ and on-site Coulomb repulsion
$U$ is coupled to a lead with a hopping matrix element $t_{1}$.
The inter-dot hopping $t_{2}$ is also present as is the inter-dot
Coulomb repulsion $V$.}
\end{figure}

The next logical step after studying individual quantum dots is to
consider systems of more than one dot. Single quantum dots are often
regarded as artificial atoms because of a similar electronic structure
and comparable number of electrons in them. By coupling several quantum
dots one is thus creating artificial molecules. Here we will not go
into detail in describing the physics of such systems. Our goal is
to compare results of our conductance formulae to results of other
methods for a double quantum dot system presented in Fig.~\ref{cap:hubsys}. 

\begin{figure}[htbp]
\begin{center}\includegraphics[%
  width=8.5cm,
  keepaspectratio]{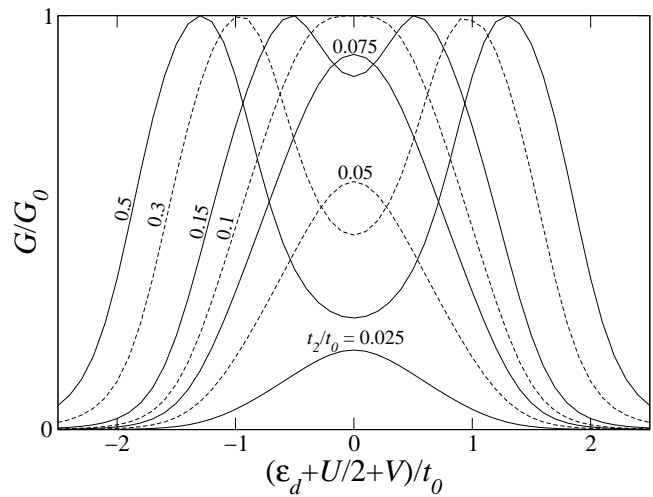}\end{center}

\caption{\label{cap:qd2ed}The zero-temperature conductance of the system
in Fig.~\ref{cap:hubsys} as a function of the position of the dot
energy level $\varepsilon_{d}$ and inter-dot hopping matrix element
$t_{2}$. The remaining parameters are $U=V=t_{0}$ and $t_{1}=0.5t_{0}$. }
\end{figure}

\begin{figure}[htbp]
\begin{center}\includegraphics[%
  width=8.5cm,
  keepaspectratio]{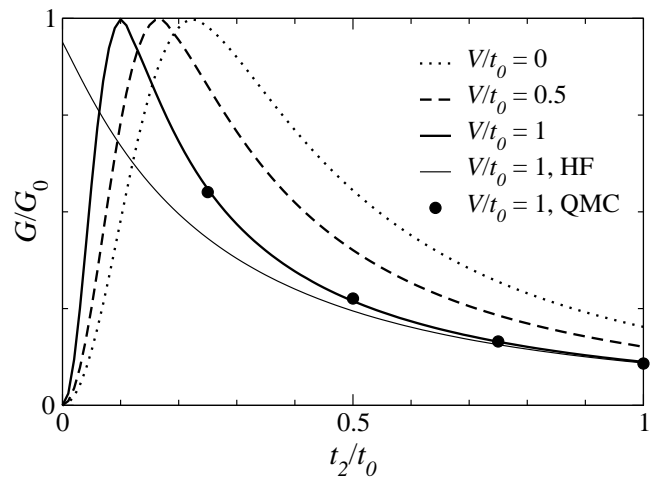}\end{center}

\caption{\label{cap:qd2res}The zero-temperature conductance of the double
quantum dot system of Fig.~\ref{cap:hubsys} at $\varepsilon_{d}+\frac{U}{2}+V=0$
as a function of the inter-dot hopping matrix element $t_{2}$ for
various values of the inter-dot Coulomb interaction $V$. As a comparison,
the Hartree-Fock and quantum Monte Carlo results \cite{Oguri97a}
are presented for $V/t_{0}=1$. Other parameters are the same as in
Fig.~\ref{cap:qd2ed}.}
\end{figure}

In the calculation we again employed the conductance formula (\ref{eq:averageS})
and calculated the ground-state energies with the variational approach
of Appendix B with the variational basis set (\ref{basis}). In Fig.~\ref{cap:qd2ed}
the zero-temperature conductance for the case where the inter-dot
and the on-site Coulomb repulsions $V$ and $U$ are of the same size,
are plotted as a function of the position of dot energy levels relative
to the Fermi energy for various values of the inter-dot hopping matrix
element $t_{2}$. The same problem in the particle-hole symmetric
case $\varepsilon_{d}+\frac{U}{2}+V=0$ was studied recently in Ref.~\cite{Oguri97a}.
The Matsubara Green's function was calculated with the quantum Monte
Carlo method and the values on discrete frequencies were extrapolated
to obtain the retarded Green's function at the Fermi energy. Then
Eqs.~(\ref{eq:FermiLB1}) and (\ref{eq:FermiLB2}) were used to calculate
the zero-temperature conductance. The results are presented in Fig.~\ref{cap:qd2res},
together with the conductance calculated within the Hartree-Fock approximation
and results of our method. The agreement with the QMC results is excellent,
while the Hartree-Fock approximation gives a qualitatively wrong conductance
curve, especially at low values of the inter-dot coupling, indicating
strong electron-electron correlations in the system. The results of
our method for lower values of $V$ are also shown in Fig.~\ref{cap:qd2res}.

\section{Summary and Conclusions}

We have demonstrated how the zero-temperature conductance of a sample
with electron-electron correlations and connected between noninteracting
leads can be determined. The method is extremely simple and is based
on several formulae connecting the conductance to persistent currents
in an auxiliary ring system. The conductance is determined only from
the ground-state energy of an interacting system, while in more traditional
approaches, one needs to know the Green's function of the system.
The Green's function approaches are often much more general, allowing
the treatment of transport at finite temperatures and for a finite
source-drain voltage applied across the sample, which in our method
is not possible. However, the advantage of the present method comes
from the fact that the ground-state energy is often relatively simple
to obtain by various numerical approaches, including variational methods,
and could therefore, for zero-temperature problems, be more appropriate.

Let us summarize the key points of the method:

(1) The {}``open'' problem of the conductance through a sample coupled
to semiinfinite leads is mapped on to a ''closed'' problem, namely
a ring threaded by a magnetic flux and containing the same correlated
electron region. 

(2) For a noninteracting sample, it is shown that the zero-temperature
conductance can be deduced from the variation of the energy of the
single-electron level at the Fermi energy with the flux in a large,
but finite ring system. The conductance is given with Eq.~(\ref{main}),
or with three simple formulae Eq.~(\ref{average}), Eq.~(\ref{halfpi})
and Eq.~(\ref{zeropi}). 

(3) Alternatively, the conductance of a noninteracting system is expressed
in terms of the variation of the ground-state energy with flux, Eq.~(\ref{total}).
Three additional conductance formulae, Eq.~(\ref{eq:averageS}),
Eq.~(\ref{eq:halfpiS}) and Eq.~(\ref{eq:drude}), are derived. 

(4) The method is primarily applicable to correlated systems exhibiting
Fermi liquid properties at zero temperature. In order to prove the
validity of the method for such systems, the concept of Fermi liquid
quasiparticles is extended to finite, but large systems. The conductance
formulae give the conductance of a system of noninteracting quasiparticles,
which is equal to the conductance of the original interacting system.
We also proved that for such systems, the ground-state energy of a
large ring system is a universal function of the magnetic flux, with
the conductance being the only parameter {[}Eqs.~(\ref{eq:even})
and (\ref{eq:odd}){]}. 

(5) The results of our method are compared to results of other approaches
for problems such as the transport through single and double quantum
dots containing interacting electrons. The comparison shows an excellent
quantitative agreement with exact Bethe ansatz results in the single
quantum dot case. The results for a double quantum dot system also
perfectly match QMC results of Ref.~\cite{Oguri97a}.

(6) One should additionally point out that in the derivation presented
in this paper we assumed the interaction in the leads to be absent.
It is clear that this assumption is not justified for all systems.
The method cannot be directly applied to systems where the interaction
in the leads is essential, as are e.g. systems exhibiting Luttinger
liquid properties. 

(7) The validity of the method is not limited to systems that do not
break the time-reversal symmetry. A generalization to systems with
a broken time-reversal symmetry, such as Aharonov-Bohm rings coupled
to leads, is possible and will be presented elsewhere~\cite{Rejec03}.

(8) Another important limitation of the present method is the single
channel approximation for the leads. It might be possible to extend
the applicability of the method to systems with multi-channel leads
by studying the influence of several magnetic fluxes that couple differently
to separate channels. This way, one might be able to probe individual
matrix elements of the scattering matrix and derive conductance formulae
relevant for such more complex systems. 

\medskip{}
\emph{Note added.} After the present work was completed the authors
met R. A. Molina and R. A. Jalabert who reported about their recent
unpublished work where an approach similar to our work is presented~\cite{Molina02,Molina03PC}.

\begin{acknowledgments}
The authors wish to acknowledge P. Prelovšek and X. Zotos for helpful
discussions and for drawing our attention to the problem of persistent
currents in correlated systems. We acknowledge V. Zlatić for discussions
related to perturbative treatment of the Anderson model and J. H.
Jefferson for useful remarks and the financial support of QinetiQ.
\end{acknowledgments}
\appendix

\section{Fermi liquid in a finite system}

In Sec.~\ref{sub:Step4} we based the proof of the validity of the
conductance formulae for Fermi liquid systems on the assumption that

\begin{eqnarray}
 &  & E\left(N,\phi;M+m\right)=E\left(N,\phi;M\right)+\quad\quad\nonumber \\
 &  & \quad+\sum_{i=1}^{m}\tilde{\varepsilon}\left(N,\phi;M;i\right)+\mathcal{O}\left(N^{-\frac{3}{2}}\right).\label{eq:energyassumpt}\end{eqnarray}
Here $E\left(N,\phi;M+m\right)$ and $E\left(N,\phi;M\right)$ are
the ground-state energies of an interacting $N$-site ring with flux
$\phi$, containing $M+m$ and $M$ electrons, respectively. $\tilde{\varepsilon}\left(N,\phi;M;i\right)$
is a single-electron energy of the ring quasiparticle Hamiltonian
$\tilde{H}\left(N,\phi;M\right)$ as defined in Eq.~(\ref{eq:assumptquasi}),
with the Fermi energy corresponding to $M$ electrons in the system.
The index $i$ labels successive single-electron energy levels above
the Fermi energy. We assume $m$ to be finite and $N$ approaching
the thermodynamic limit. In this Appendix we will give arguments showing
that the assumption of Eq.~(\ref{eq:energyassumpt}) is indeed valid.
In Sec.~1 we first express the problem in terms of the Green's function
of the system. In Sec.~2 we study the properties of the self-energy
due to interaction in finite ring systems and then use this results
to complete the proof in Sec.~3.

\subsection{Relation to the Green's function\label{sub:A1}}

Assume we manage to prove Eq.~(\ref{eq:energyassumpt}) for $m=1$,
i.e.\begin{eqnarray}
 &  & E\left(N,\phi;M+1\right)=E\left(N,\phi;M\right)+\nonumber \\
 &  & \quad+\tilde{\varepsilon}\left(N,\phi;M;1\right)+\mathcal{O}\left(N^{-\frac{3}{2}}\right).\label{eq:energyassumpt1}\end{eqnarray}
Then we can use the same result to relate the energy of a system with
$M+2$ electrons to that with $M+1$ electrons, \begin{eqnarray}
 &  & \!\!\!\!\!\!\!\!\!\!\!\!\!\!\!\!\!\! E\left(N,\phi;M+2\right)=E\left(N,\phi;M\right)+\nonumber \\
 &  & \!\!\!\!\!\!\!\!\!\!\!\!\!\!\!\!\!\!\quad+\tilde{\varepsilon}\left(N,\phi;M;1\right)+\tilde{\varepsilon}\left(N,\phi;M+1;1\right)+\mathcal{O}\left(N^{-\frac{3}{2}}\right).\end{eqnarray}
Now the matrix elements of quasiparticle Hamiltonians $\tilde{H}\left(N,\phi;M+1\right)$
and $\tilde{H}\left(N,\phi;M\right)$ differ by an amount of the order
of $\frac{1}{N}$. To see this, note that the shift of the Fermi energy
as an electron is added to the system is of the order of $\frac{1}{N}$,
producing a shift of the same order in the self-energy and it's derivative
at the Fermi energy, which define the quasiparticle Hamiltonian through
Eq.~(\ref{eq:assumptquasi}). As the difference of the Hamiltonians
$\Delta\tilde{H}$ is small, we can use the first order perturbation
theory,\begin{eqnarray}
 &  & \!\!\!\!\!\!\!\!\!\!\!\!\!\tilde{\varepsilon}\left(N,\phi;M+1;1\right)=\nonumber \\
 &  & \!\!\!\!\!\!\!\!\!\!\!\!\!\quad=\tilde{\varepsilon}\left(N,\phi;M;2\right)+\left\langle N,\phi;M;2\left|\Delta\tilde{H}\right|N,\phi;M;2\right\rangle =\nonumber \\
 &  & \!\!\!\!\!\!\!\!\!\!\!\!\!\quad=\tilde{\varepsilon}\left(N,\phi;M;2\right)+\mathcal{O}\left(N^{-2}\right).\label{eq:aswedid}\end{eqnarray}
In the last step we made use of the fact that the quasiparticle Hamiltonians
differ only in a finite number of sites in and in the vicinity of
the central region, and of the fact that the amplitude of the quasiparticle
single-electron wavefunction $\left|N,\phi;M;2\right\rangle $ is
of the order of $\frac{1}{\sqrt{N}}$. Thus we have proved Eq.~(\ref{eq:energyassumpt})
for $m=2$ and using the same procedure, we can extend the proof to
any finite $m$.

To complete the proof, we still need to show the validity of Eq.~(\ref{eq:energyassumpt1}).
As a first step, consider the Lehmann representation of the zero-temperature
central region Green's function \begin{equation}
G_{ji}\left(t,t^{\prime}\right)=-i\theta\left(t-t^{\prime}\right)\bigl<0\bigl|\bigl[d_{j}\left(t\right),d_{i}^{\dagger}\left(t^{\prime}\right)\bigr]\bigr|0\bigr>\end{equation}
 of a ring system characterized with $N$ and $\phi$, containing
$M$ electrons,\begin{eqnarray}
 &  & G_{ji}\left(N,\phi;M;z\right)=\nonumber \\
 &  & \quad=\quad\sum_{n}\frac{\bigl<0\bigr|d_{j}\bigl|n\bigr>\bigl<n\bigr|d_{i}^{\dagger}\bigl|0\bigr>}{z-\left(E_{n}^{M+1}-E_{0}^{M}\right)}+\nonumber \\
 &  & \quad\quad+\sum_{n}\frac{\bigl<0\bigr|d_{i}^{\dagger}\bigl|n\bigr>\bigl<n\bigr|d_{j}\bigl|0\bigr>}{z-\left(E_{0}^{M}-E_{n}^{M-1}\right)}.\label{eq:Lehmann}\end{eqnarray}
The first sum runs over all basis states with $M+1$ electrons, while
the second sum runs over the states with $M-1$ electrons. The difference
in the ground-state energies of systems with $M+1$ and $M$ electrons
is evidently equal to the position of the first $\delta$-peak above
the Fermi energy in the spectral density corresponding to the Green's
function. In what follows, we will try to determine the energy of
this $\delta$-peak.

\subsection{Self-energy due to interaction\label{sub:A2}}

To achieve the goal we have set in the previous Section, we first
need to study the structure of the self-energy due to interaction
in a finite ring with flux. Let us again consider the Lehmann representation
(\ref{eq:Lehmann}) and to be specific, limit ourselves to states
above the Fermi energy. Introducing $\varphi_{j}^{n}=\left\langle 0\left|d_{j}\right|n\right\rangle $
and $\varepsilon_{n}=E_{n}^{M+1}-E_{n}^{M}$, we can express the Green's
function as\begin{equation}
G_{ji}\left(N,\phi;M;z\right)=\sum_{n}\frac{\varphi_{j}^{n}\varphi_{i}^{n\ast}}{z-\varepsilon_{n}}.\end{equation}
 This expression can also be interpreted as a local Green's function
of a larger \emph{noninteracting} system, consisting of the central
region and a bath of noninteracting energy levels, the number of which
is equal to the number of multi-electron states with $M+1$ and $M-1$
electrons of the original interacting system. The self-energy due
to {}``hopping out of the central region'', which includes both
the effects of the interaction as well as those due to the hopping
into the ring, can then be expressed as \begin{equation}
\Sigma_{ji}\left(N,\phi;M;z\right)=\sum_{n}\frac{V_{jn}V_{ni}}{z-\varepsilon_{n}},\label{eq:selffinite}\end{equation}
where $V_{jn}$ are the ''hopping matrix elements'' between the
central region and the {}``bath''. Thus we have shown that, as far
as the single-electron Green's function is concerned, the interacting
system can be mapped on a larger, but noninteracting system. 

\begin{figure}[htbp]
\begin{center}\begin{tabular}{cc}
\includegraphics[%
  bb=140bp 560bp 325bp 670bp,
  clip,
  width=4cm,
  keepaspectratio]{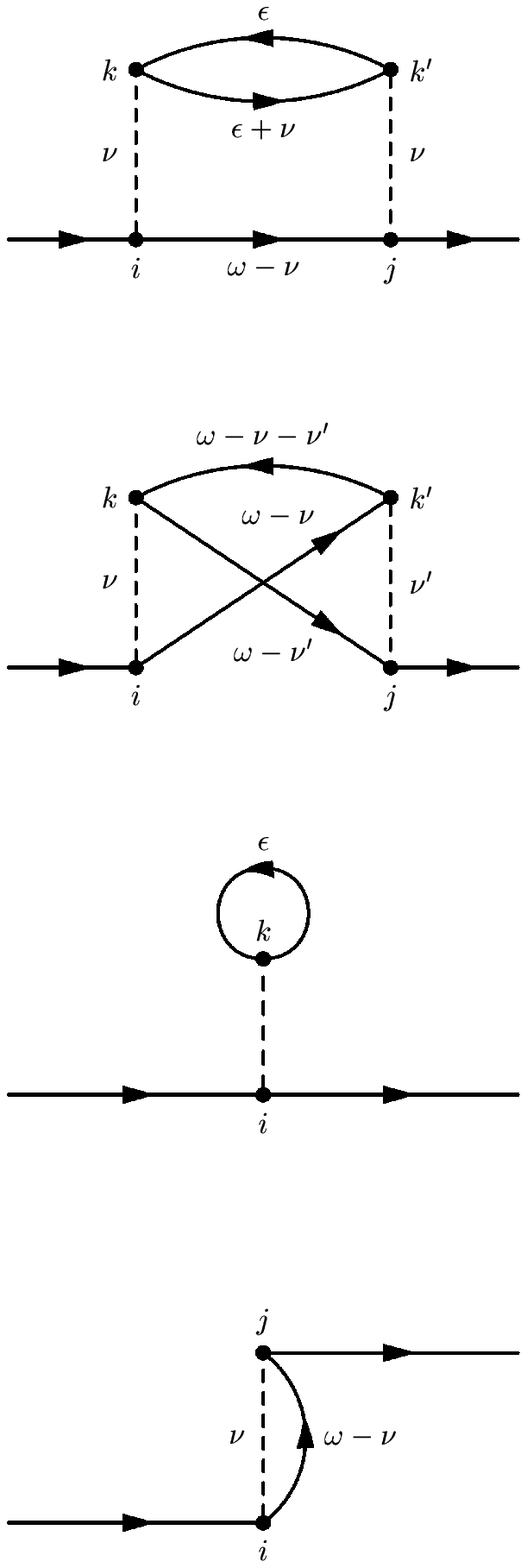}&
\includegraphics[%
  bb=140bp 418bp 325bp 523bp,
  clip,
  width=4cm,
  keepaspectratio]{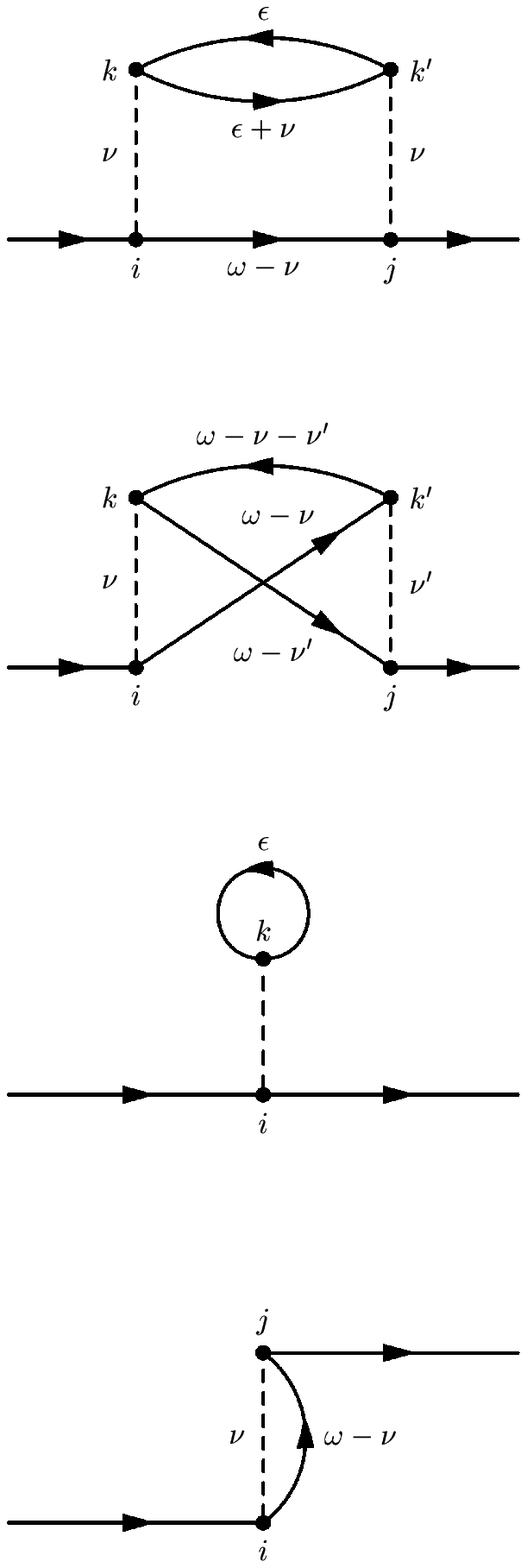}\tabularnewline
\end{tabular}\end{center}

\caption{Second-order self-energy diagrams.\label{cap:Feyn2}}
\end{figure}

\begin{figure}[htbp]
\begin{center}\includegraphics[%
  height=8.5cm,
  keepaspectratio,
  angle=-90,
  origin=lB]{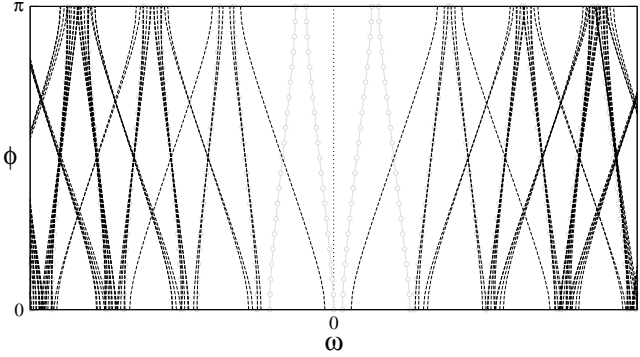}\end{center}

\caption{\label{cap:self}Dashed lines show the positions of $\delta$-peaks
(\ref{eq:selffinite2}) in the second-order self-energy corresponding
to single-electron energy levels of an unperturbed system presented
with gray lines. }
\end{figure}

To further clarify the concepts introduced above, we calculated the
self-energy due to interaction within the second-order perturbation
theory. Following the calculations by Horvatić, Šokčević and Zlatić
\cite{Horvatic80,Horvatic85,Horvatic87} for the Anderson model, we
sum the second order self-energy diagrams shown in Fig.~\ref{cap:Feyn2},
including Hartree and Fock terms into the unperturbed Hamiltonian.
A lengthy but straightforward calculation, which we do not repeat
here, shows that one can identify the states $n$ of Eq.~(\ref{eq:selffinite})
with three Hartree-Fock single-electron state indices $\mathbf{q}=\left(q_{1},q_{2},q_{3}\right)$
such that $q_{1}$ and $q_{2}$ are above the Fermi energy and $q_{3}$
is below it (or vice versa), and a spin index $s$. The {}``bath''
energy levels \begin{equation}
\varepsilon_{\mathbf{q}s}=\varepsilon_{q_{1}}+\varepsilon_{q_{2}}-\varepsilon_{q_{3}}\label{eq:selffinite2}\end{equation}
and the {}``hopping matrix elements'' related to the self-energy
for electrons with spin $\sigma$\begin{equation}
V_{j\mathbf{q}s}=\left\{ \begin{array}{ll}
\sum_{k^{\prime}}U_{jk^{\prime}}^{\sigma\bar{\sigma}}\varphi_{j}^{q_{1}}\varphi_{k^{\prime}}^{q_{2}}\varphi_{k^{\prime}}^{q_{3}\ast}, & s=\bar{\sigma},\\
\frac{1}{\sqrt{2}}\sum_{k^{\prime}}U_{jk^{\prime}}^{\sigma\sigma}\left[\varphi_{j}^{q_{1}}\varphi_{k^{\prime}}^{q_{2}}-\varphi_{j}^{q_{2}}\varphi_{k^{\prime}}^{q_{1}}\right]\varphi_{k^{\prime}}^{q_{3}\ast}, & s=\sigma\end{array}\right.\end{equation}
are then expressed in terms of the Coulomb interaction matrix elements
(\ref{eq:HU}), and the Hartree-Fock single-electron energies $\varepsilon_{q}\left(N,\phi;M\right)$
and the corresponding wavefunctions $\left|\varphi^{q}\left(N,\phi;M\right)\right\rangle $.
In Fig.~\ref{cap:self} the positions of $\delta$-peaks in the imaginary
part of the self-energy as a function of magnetic flux through the
ring are plotted. Note that as the flux is varied, the positions of
the peaks fluctuate by an amount of the order of the single-electron
level spacing which is of the order of $\frac{1}{N}$. The weights
of the peaks also depend on the flux. A similar behavior is expected
if higher order processes are also taken into account. 


Finally, let us study the self-energy in the thermodynamic limit.
We will show that in this case, the self-energy is independent of
flux and is equal to the self-energy of the original, two-lead system,
shown in Fig.~\ref{cap:System}. To prove this statement, we consider
a self-energy Feynman diagram for the central region decoupled from
the ring, which then is obviously independent of flux. To calculate
the self-energy for the full system, one should insert the self-energy
due to hopping into the ring into each propagator of the diagram.
The self-energy due to hopping into the ring is\begin{equation}
\Sigma_{ji}^{\left(0\right)}\left(N,\phi;z\right)=\sum_{k}\frac{V_{jk}V_{ki}}{z-\varepsilon_{k}},\label{eq:self0LR}\end{equation}
 where $\varepsilon_{k}$ are the single-electron energy levels of
the ring decoupled from the central region and $V_{ki}=-\psi_{L}^{k}t_{Li}-\psi_{R}^{k}t_{Ri}$
is the hopping matrix element between site $i$ in the central region
and the single-electron state $k$ in the ring. $V_{ki}$ is expressed
in terms of the hopping matrix element $t_{Li}$ between the site
$i$ and the ring site $L$ adjacent to the central region and the
single-electron wavefunction $\psi_{L}^{k}={\scriptstyle \sqrt{\frac{2}{N+1}}}\sin k$
at site $L$, where $N$ is the number of sites in the ring. There
is also a similar contribution to $V_{ki}$ corresponding to the hopping
into the right lead. In the ring system, the right lead wavefunction
can be expressed in terms of the left lead one as $\psi_{R}^{k}=\left(-1\right)^{n}e^{-i\phi}\psi_{L}^{k}$
with $k=\frac{n\pi}{N+1}$, if one takes into account the parity of
the wavefunctions and the effect of the flux. Thus, Eq.~(\ref{eq:self0LR})
transforms into\begin{eqnarray}
 &  & \!\!\!\!\!\!\!\!\!\!\!\!\!\!\!\!\!\!\!\!\Sigma_{ji}^{\left(0\right)}\left(N,\phi;z\right)=\Sigma_{ji}^{\left(L\right)}\left(N;z\right)+\Sigma_{ji}^{\left(R\right)}\left(N;z\right)+\nonumber \\
 &  & \!\!\!\!\!\!\!\!\!\!\!\!\!\!\!\!\!\!\!\!\quad+\frac{2\left(t_{jL}t_{Ri}e^{-i\phi}+t_{jR}t_{Li}e^{i\phi}\right)}{N+1}\sum_{k}\frac{\left(-1\right)^{n}\sin^{2}k}{z-\varepsilon_{k}},\end{eqnarray}
 where $\Sigma_{ji}^{\left(L\right)}\left(N;z\right)$ and $\Sigma_{ji}^{\left(R\right)}\left(N;z\right)$
are the self-energies due to hopping into the left and the right leads
(each with $N$ sites) of the two-lead system. In the third term,
one can perform the sum over odd $n$-s and over even $n$-s separately.
The sums differ only in sign in the $N\rightarrow\infty$ limit and
therefore, this term vanishes. Therefore, in the thermodynamic limit
the self-energy due to interaction is the same in both two-lead and
ring systems.

\subsection{Proof of Eq.~(\ref{eq:energyassumpt1})\label{sub:A3}}

Positions of $\delta$-peaks in the spectral density of the interacting
system correspond to the single-electron energy levels of the noninteracting
part of the ring Hamiltonian coupled to the {}``bath'' according
to Eq.~(\ref{eq:selffinite}). These energies can be obtained by
solving for zeroes of the determinant of the inverse of the {}``local''
Green's function\begin{equation}
\det\left[\omega\mathbf{1}-\mathbf{H}^{\left(0\right)}\left(N,\phi\right)-\bm\Sigma\left(N,\phi;M;\omega+i\delta\right)\right]=0.\label{eq:zeroes}\end{equation}
 What we are going to prove in this Section is that the lowest positive
solution of this equation corresponds to $\tilde{\varepsilon}\left(N,\phi;M;1\right)$
as required by Eq.~(\ref{eq:energyassumpt1}). 

\begin{figure}[htbp]
\begin{center}\includegraphics[%
  width=8.5cm,
  keepaspectratio]{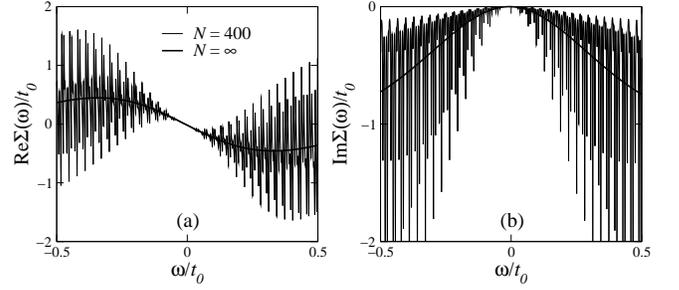}\end{center}

\caption{\label{cap:selfcomp}The (a) real and (b) imaginary parts of the
self-energy of an interacting system in the thermodynamic limit and
for $N=400$ with $\phi=\frac{3\pi}{4}$. The system is described
in Fig.~\ref{cap:sigma}.}
\end{figure}

We begin by separating the self-energy at frequencies close to the
Fermi energy into two contributions, one ($\bm\Sigma^{\prime\prime}$)
due to the {}``bath'' states close to the Fermi energy and the other
($\bm\Sigma^{\prime}$) of all the other states with energies which
are separated from the expected solution of Eq.~(\ref{eq:zeroes})
by at least an amount of the order of the single-electron level spacing
$\Delta$, which is of the order of $\frac{1}{N}$. We first estimate
the second term. Let us divide the frequency axis into intervals of
width $\Delta$, each contributing to the self-energy at $\left|\omega\right|<\Delta$
an amount given by\begin{equation}
\int_{\varepsilon}^{\varepsilon+\Delta}\frac{\bm\rho\left(\varepsilon\right)}{\omega-\varepsilon}\mathrm{d}\varepsilon,\end{equation}
 where $\rho_{ji}\left(\varepsilon\right)=\sum_{n}V_{jn}V_{ni}\delta\left(\varepsilon-\varepsilon_{n}\right)$
if the notation of Eq.~(\ref{eq:selffinite}) is used. On average,
this contribution corresponds to that of a system in the thermodynamic
limit where $\bm\rho\left(\varepsilon\right)$ is a continuous function
and the magnitude of each contribution is at most of the order of
$\frac{1}{N}$. To see this, let us assume $\bm\rho\left(\varepsilon\right)$
is proportional to $\varepsilon^{2}$ (\ref{Luttinger}) for all values
of $\varepsilon$ up to a cutoff of the order of 1. Such an approximation
can be considered as the upper limit of possible values of $\bm\rho\left(\varepsilon\right)$
in Fermi liquid systems, if one does not take into account the rapidly
decreasing tails at higher energies, which contribute a negligible
amount to the self-energy at the Fermi energy. Evaluating the above
integral, we find that contributions of the intervals close to the
Fermi energy are of the order of $\frac{1}{N^{2}}$ and contributions
of the intervals near the cutoff are of the order of $\frac{1}{N}$.
Using an analogous procedure, we can also evaluate the derivative
of the self-energy close to the Fermi energy, with contributions \begin{equation}
-\int_{\varepsilon}^{\varepsilon+\Delta}\frac{\bm\rho\left(\varepsilon\right)}{\left(\omega-\varepsilon\right)^{2}}\mathrm{d}\varepsilon.\end{equation}
 In this case, also contributions corresponding to intervals close
to the Fermi energy are of the order of $\frac{1}{N}$. If $\bm\rho\left(\varepsilon\right)$
for a finite $N$ is used instead, there are large fluctuations about
the average value (see the discussion in the previous Section) with
the amplitude of fluctuations of the same order of magnitude as the
average value itself. To estimate the difference between the finite-system's
real part of the self-energy (or its derivative) close to the Fermi
energy and the corresponding quantity for a system in the thermodynamic
limit, we note that a sum of $N$ quantities, each of them of the
order of $\frac{1}{N}$ with a standard deviation of the same order
of magnitude, has a standard deviation of the order of $N^{-\frac{1}{2}}$,
and therefore, we can estimate that for $\left|\omega\right|<\Delta$\begin{eqnarray}
 &  & \bm\Sigma^{\prime}\left(N,\phi;M;\omega+i\delta\right)=\nonumber \\
 &  & \quad=\bm\Sigma\left(0+i\delta\right)+\mathcal{O}\left(N^{-\frac{1}{2}}\right),\\
 &  & \left.\frac{\partial\bm\Sigma^{\prime}\left(N,\phi;M;\omega+i\delta\right)}{\partial\omega}\right|_{\omega}=\nonumber \\
 &  & \quad=\left.\frac{\partial\bm\Sigma\left(\omega+i\delta\right)}{\partial\omega}\right|_{\omega=0}+\mathcal{O}\left(N^{-\frac{1}{2}}\right).\end{eqnarray}
Note that we do not need to exclude the contribution of the interval
at the Fermi energy (the one corresponding to $\bm\Sigma^{\prime\prime}$)
from self-energies in the right hand sides of these equations, because
the corresponding contributions are smaller than $N^{-\frac{1}{2}}$
as discussed above. Also the errors arising from the fact that the
right hand sides are evaluated at $\omega=0$ instead of at $\omega$
are only of the order of $\frac{1}{N}$, as discussed in the previous
Section. In Fig.~\ref{cap:selfcomp} a comparison of the self-energies
at a finite $N$ and in the thermodynamic limit is presented. Note
that in the vicinity of the Fermi energy, the real parts of both self-energies
coincide. 

One can now proceed as in Eqs.~(\ref{eq:zzzz}) and (\ref{quasiham}),
defining the renormalization matrix $\mathbf{Z}^{\prime}\left(N,\phi;M\right)=\mathbf{Z}+\mathcal{O}\bigl(N^{-\frac{1}{2}}\bigr)$
and the quasiparticle Hamiltonian $\tilde{\mathbf{H}}^{\prime}\left(N,\phi;M\right)=\tilde{\mathbf{H}}\left(N,\phi;M\right)+\mathcal{O}\bigl(N^{-\frac{1}{2}}\bigr)$
corresponding to the self-energy $\bm\Sigma^{\prime}$. As shown in
the previous Section, the self-energies of an infinite two-lead system
and the corresponding ring system are the same and therefore, the
renormalized matrix elements of $\tilde{\mathbf{H}}\left(N,\phi;M\right)$
correspond to those of a two-lead system. For $\left|\omega\right|<\Delta$,
Eq.~(\ref{eq:zeroes}) transforms into

\begin{equation}
\det\left[\omega\mathbf{1}-\tilde{\mathbf{H}}^{\prime}\left(N,\phi;M\right)-\tilde{\bm\Sigma}^{\prime\prime}\left(N,\phi;M;\omega+i\delta\right)\right]=0,\label{eq:zeroes1}\end{equation}
 where the coupling to the remaining {}``bath'' levels has been
renormalized as $\tilde{\bm\Sigma}^{\prime\prime}=\mathbf{Z}^{\prime1/2}\bm\Sigma^{\prime\prime}\mathbf{Z}^{\prime1/2}$.
Let us for a moment neglect this term in Eq.~(\ref{eq:zeroes1}).
As the difference $\Delta\tilde{H}$ between Hamiltonians $\tilde{H}$$\left(N,\phi;M\right)$
and $\tilde{H}^{\prime}\left(N,\phi;M\right)$ is small for a large
$N$, one is justified to relate their single-electron energy levels
using the first order perturbation formula\begin{eqnarray}
 &  & \tilde{\varepsilon}^{\prime}\left(N,\phi;M;1\right)=\nonumber \\
 &  & \quad=\tilde{\varepsilon}\left(N,\phi;M;1\right)+\left\langle N,\phi;M;1\left|\Delta\tilde{H}\right|N,\phi;M;1\right\rangle =\nonumber \\
 &  & \quad=\tilde{\varepsilon}\left(N,\phi;M;1\right)+\mathcal{O}\left(N^{-\frac{3}{2}}\right).\label{eq:epsprime}\end{eqnarray}
 In the last step we made use of arguments similar to those in deriving
Eq.~(\ref{eq:aswedid}).

The energy (\ref{eq:epsprime}) can acquire an additional shift because
of the coupling $\bm\Sigma^{\prime\prime}$. To estimate this shift
we first note that in the worst case scenario, i.e. when there is
a single bath energy level which coincides with the quasiparticle
energy level (\ref{eq:epsprime}), the coupling matrix elements $V_{jn}$
in Eq.~(\ref{eq:selffinite}) must be at most of the order of $N^{-\frac{3}{2}}$
for Eq.~(\ref{Luttinger}) to be satisfied in the thermodynamic limit.
Then one can make use of the degenerate first order perturbation theory,
which shows that the quasiparticle energy level is shifted by an additional
amount of the order of $\frac{1}{N^{2}}$. This completes the proof
of Eq.~(\ref{eq:energyassumpt}).

\begin{figure}[htbp]
\begin{center}\includegraphics[%
  bb=140bp 3bp 495bp 365bp,
  clip,
  width=8.5cm,
  keepaspectratio]{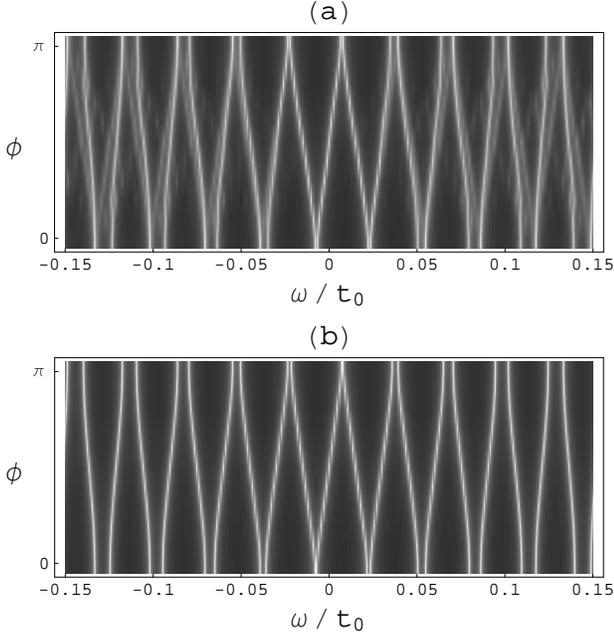}\end{center}

\caption{\label{cap:finitefl2}(a) The total density of states of an interacting
ring system within the second-order perturbation theory. (b) Total
density of states corresponding to the quasiparticle Hamiltonian.
The system is described in Fig.~\ref{cap:sigma}.}
\end{figure}

As a conclusion, in Fig.~\ref{cap:finitefl2} we present a comparison
of the total densities of states for a finite ring interacting system
within the second order perturbation theory and in the quasiparticle
Hamiltonian approximation. Note that the states near the Fermi energy
are well described with the quasiparticle approximation, while the
states further away from the Fermi energy are split in the interacting
case. Similar results were reported in Ref.~\cite{Altshuler97}.

\section{Variational ground-state energy}

In order to calculate the conductance for interacting systems, we
first need to devise a robust method that would allow us to efficiently
calculate the ground-state energy of such systems. Note that we need
a method that would provide us with the energy of a system with a
very large number of sites in the ring. However, the number should
still be finite - i.e. we must not perform the calculations in the
thermodynamic limit. We made use of the projection method of Gunnarson
and Schönhammer \cite{Schonhammer75,Schonhammer76,Gunnarson85}, introduced
originally to calculate the ground-state energy of the Anderson impurity
model, and extended it to treat the more general Hamiltonian (\ref{eq:Hamiltonian}).

Let us introduce an auxiliary noninteracting Hamiltonian,\begin{equation}
\tilde{H}=H_{L}+\tilde{V_{L}}+\tilde{H}_{C}^{\left(0\right)}+\tilde{V}_{R}+H_{R},\label{auxillary}\end{equation}
 with arbitrary matrix elements describing the hopping between the
leads and the central region, and the central region itself. Note
that these are the same matrix elements as the ones being renormalized
in the Fermi liquid quasiparticle Hamiltonian (\ref{quasiham}). Let
us also define a Hilbert space spanned by a set of $4^{M}$ basis
functions\begin{equation}
\left|\psi_{\bm\alpha}\right\rangle \equiv P_{\bm\alpha}\left|\tilde{0}\right\rangle \equiv\prod_{i\in C}P_{\alpha_{i}}^{i}\left|\tilde{0}\right\rangle ,\label{basis}\end{equation}
 where $M$ is the number of sites in the central region, $\left|\tilde{0}\right\rangle $
is the ground-state of the auxiliary Hamiltonian (\ref{auxillary})
containing the same number of electrons as there are in the ground
state of the original Hamiltonian, and \begin{eqnarray}
P_{0}^{i} & = & \left(1-n_{i\uparrow}\right)\left(1-n_{i\downarrow}\right),\\
P_{\sigma}^{i} & = & n_{i\sigma}\left(1-n_{i\bar{\sigma}}\right),\\
P_{2}^{i} & = & n_{i\uparrow}n_{i\downarrow}\end{eqnarray}
 are projection operators on unoccupied, singly occupied and doubly
occupied site $i$. The original Hamiltonian is diagonalized in the
reduced basis set introduced above, \begin{equation}
H_{\bm\beta\bm\alpha}=ES_{\bm\beta\bm\alpha},\label{eq:evprob}\end{equation}
 with $H_{\bm\beta\bm\alpha}=\left\langle \psi_{\bm\beta}\left|H\right|\psi_{\bm\alpha}\right\rangle $
being the matrix elements of the Hamiltonian and $S_{\bm\beta\bm\alpha}=\left\langle \psi_{\bm\beta}\mid\psi_{\bm\alpha}\right\rangle $
take into account the fact that the basis functions do not form an
orthonormal basis set. The eigenstate with the lowest energy $E_{\tilde{H}}$
of this eigenvalue problem is an approximation to the ground-state
energy of the original Hamiltonian. Varying the parameters of the
auxiliary Hamiltonian, one can find their optimal values which minimize
$E_{\tilde{H}}$. The solution of this minimization problem is the
final approximation to the ground-state energy.

Let us consider some simple limits of the problem. In the noninteracting
case where $U=0$, one can choose the auxiliary Hamiltonian to be
equal to the true Hamiltonian $\tilde{H}=H$. Then the wavefunction
$\left|\psi\right\rangle =\sum_{\bm\alpha}P_{\bm\alpha}\left|\tilde{0}\right\rangle =\left|\tilde{0}\right\rangle =\left|0\right\rangle $
is the exact ground-state wavefunction of the system. Note that applying
the same wavefunction ansatz to the interacting case and allowing
the matrix elements of the auxiliary Hamiltonian to be renormalized,
provides us with the Hartree-Fock solution of the problem. Therefore,
the variational method introduced above always gives the ground-state
energy which is lower or equal to the corresponding Hartree-Fock ground-state
energy. In the limit of the central region being decoupled from the
ring, i.e. $V_{L}=V_{R}=0$, the variational method also yields the
exact ground-state energy. To prove this statement, let us select
the matrix elements of $\tilde{H}$ in such a way that in its ground
state there are $m$ electrons in the central region. Then the basis
set (\ref{basis}) spans the full Hilbert space for $m$ electrons
in the central region. As there is no coupling to the states in the
ring, solving the eigenvalue problem (\ref{eq:evprob}) provides us
with the exact ground state of the problem with a constraint of a
fixed number of electrons in the central region. By varying $\tilde{H}$,
all the possible values of $m$ can be tested and the one yielding
the lowest ground-state energy corresponds to the correct ground state
of the system.

The variational wavefunction ansatz can be improved by extending the
Hilbert space with additional basis functions, the most promising
candidates being of type\cite{Gunnarson85} \begin{equation}
\left|\psi_{\bm\beta\lambda ji\sigma\bm\alpha}\right\rangle =P_{\bm\beta}\hat{V}_{\lambda ji\sigma}P_{\bm\alpha}\left|\tilde{0}\right\rangle ,\label{eq:extbas}\end{equation}
 where $\hat{V}_{\lambda ji\sigma}=V_{\lambda ji}c_{j\sigma}^{\dagger}d_{i\sigma}+\mathrm{h}.\mathrm{c}.$
and $\lambda$ is a lead index, i.e. either $L$ or $R$. On the other
hand, as the size of the Hilbert space increases exponentially with
the number of sites in the central region, it might be convenient
to limit the basis set to the states obtained by projecting to the
central region's many body states between which fluctuations are possible.

Finally, we state some technical details concerning the evaluation
of $H_{\bm\beta\bm\alpha}$ and $S_{\bm\beta\bm\alpha}$. It is convenient
to express these matrix elements only in terms of quantities related
to the central region and the neighboring sites in the leads. As \begin{equation}
S_{\bm\beta\bm\alpha}=\left\langle \tilde{0}\left|P_{\bm\beta}P_{\bm\alpha}\right|\tilde{0}\right\rangle =\left\langle \tilde{0}\left|P_{\bm\alpha}\right|\tilde{0}\right\rangle \delta_{\bm\beta\bm\alpha},\label{sab}\end{equation}
 the scalar products between the basis functions are evidently expressed
with the central region quantities. The matrix elements of the Hamiltonian
can be expressed as \begin{eqnarray}
H_{\bm\beta\alpha} & = & \left\langle \tilde{0}\left|P_{\bm\beta}HP_{\bm\alpha}\right|\tilde{0}\right\rangle =\nonumber \\
 & = & \left\langle \tilde{0}\left|\tilde{H}P_{\bm\beta}P_{\bm\alpha}\right|\tilde{0}\right\rangle +\left\langle \tilde{0}\left|P_{\bm\beta}HP_{\bm\alpha}\right|\tilde{0}\right\rangle -\nonumber \\
 &  & -\left\langle \tilde{0}\left|\tilde{H}P_{\bm\beta}P_{\bm\alpha}\right|\tilde{0}\right\rangle =\nonumber \\
 & = & \tilde{E}S_{\bm\beta\bm\alpha}+\left\langle \tilde{0}\left|P_{\bm\beta}\left(V_{L}+H_{\mathrm{C}}+V_{R}\right)P_{\bm\alpha}\right|\tilde{0}\right\rangle -\nonumber \\
 &  & -\left\langle \tilde{0}\left|\left(\tilde{V}_{L}+\tilde{H}_{\mathrm{C}}^{\left(0\right)}+\tilde{V}_{R}\right)P_{\bm\alpha}\right|\tilde{0}\right\rangle \delta_{\bm\beta\bm\alpha},\quad\quad\label{eq:hab}\end{eqnarray}
 where $\tilde{E}$ is the ground-state energy of the auxiliary Hamiltonian
$\tilde{H}$. In the second and the third term we made use of the
fact that lead Hamiltonians $H_{\mathrm{L}}$ and $H_{R}$ commute
with the central region projectors and therefore, they cancel out.
Again, we succeeded in expressing the matrix elements in terms of
central region quantities together with quantities related to the
neighboring sites in the leads. Similar results are obtained if the
extended basis set of Eq.~(\ref{eq:extbas}) is used. The matrix
elements in Eqs.~(\ref{sab}) and (\ref{eq:hab}) need to be calculated
in a noninteracting state. Therefore, we can make use of the Wick's
theorem to decompose the expressions into two-operator averages of
type $\left\langle \tilde{0}\left|d_{j\sigma}^{\dagger}d_{i\sigma}\right|\tilde{0}\right\rangle $.
As a huge number of terms is generated in this procedure, the decomposition
was performed automatically by symbolic manipulation of operators.
The ground-state energy of the auxiliary Hamiltonian and the two-operator
averages can be expressed in terms of the single-electron energies
$\tilde{\varepsilon}_{k}$ and wavefunctions $\left|\tilde{\varphi}^{k}\right\rangle $
of $\tilde{H}$ as\begin{eqnarray}
\tilde{E} & = & 2\sum_{k\,\mathrm{occ}.}\tilde{\varepsilon}_{k},\\
\left\langle \tilde{0}\left|d_{j\sigma}^{\dagger}d_{i\sigma}\right|\tilde{0}\right\rangle  & = & \sum_{k\,\mathrm{occ}.}\tilde{\varphi}_{j}^{k\ast}\tilde{\varphi}_{i}^{k}.\end{eqnarray}
 The sums run only over the single-electron states occupied in the
ground state $\left|\tilde{0}\right\rangle $. The eigenvalues $\tilde{\varepsilon}_{k}$
were calculated in a basis in which the Hamiltonian matrix is banded,
i.e. linear combinations of local basis functions corresponding to
the left lead and right lead sites were introduced to {}``move''
the hopping matrix elements in corners of the matrix close to the
diagonal. For each eigenvalue, only the components of the eigenvector
related to the central region and neighboring sites were calculated,
again taking the special structure of the matrix into account. The
procedure used scales with the number of sites in the ring as $\mathcal{O}\left(N^{2}\right)$,
which allows one to treat systems with up to 10000 sites in the ring.

\bibliographystyle{prsty}
\bibliography{thesis}

\end{document}